\documentclass{article}

\usepackage{arxiv}

\usepackage{algorithm}
\usepackage{algpseudocode}
\usepackage[utf8]{inputenc} 
\usepackage[T1]{fontenc}    
\usepackage{hyperref}       
\usepackage{url}            
\usepackage{booktabs}       
\usepackage{amsfonts}       
\usepackage{nicefrac}       
\usepackage{microtype}      
\usepackage{lipsum}
\usepackage{fancyhdr}       
\usepackage{graphicx}       
\graphicspath{{media/}}     
\usepackage{amsmath}
\usepackage{float}
\usepackage[utf8]{inputenc}
\usepackage{subcaption} 
\usepackage{multirow}%
\usepackage{multicol}
\usepackage{placeins}
\usepackage{cite}

\title{SPECTRA: A Physics-informed Digital Twin for Real-time Structural Anomaly Inference under Operational Variability}

\author{
	 \href{https://orcid.org/0009-0004-4945-9821}{\includegraphics[scale=0.06]{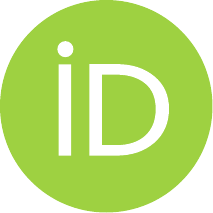}\hspace{1mm}Anshu Sharma}\\
	Department of Civil \& Environmental Engineering,\\
		University of Strathclyde,\\
	Glasgow, UK \\
	\texttt{anshu.sharma@strath.ac.uk} \\
	\And
	\href{https://orcid.org/0000-0001-7782-513X}{\includegraphics[scale=0.06]{orcid.pdf}\hspace{1mm}Basuraj Bhowmik*}\thanks{Corresponding author}\\
	Department of Civil \& Environmental Engineering,\\
	University of Strathclyde,\\
	Glasgow, UK \\
	\texttt{basuraj.bhowmik@strath.ac.uk} \\
}

\renewcommand{\shorttitle}{SPECTRA}

\hypersetup{
pdftitle={A template for the arxiv style},
pdfsubject={q-bio.NC, q-bio.QM},
pdfauthor={David S.~Hippocampus, Elias D.~Striatum},
pdfkeywords={First keyword, Second keyword, More},
}

\begin{document}
\maketitle

\begin{abstract} 
	Structural health monitoring is moving from damage detection alone towards real-time decision support for ageing and safety-critical infrastructure. This shift requires monitoring methods that can separate true structural change from benign environmental and operational variability, while remaining interpretable to engineers. This paper presents SPECTRA, a physics-informed eigen-compressed digital twin framework for real-time structural anomaly inference. The framework combines a healthy structural twin, eigen-compressed dynamic representation, full-order twin innovation, residual-augmented spectral features, kernel principal component analysis, and a persistent decision rule. The central idea is that structural
	anomalies are inferred not from statistical features alone, but from disagreement between the
	measured response and a physics-informed healthy twin. The method is assessed through seven numerical benchmarks: smooth Duffing-type nonlinear drift, sudden stiffness loss, gradual stiffness degradation, bilinear breathing stiffness, environmental and operational variability-confounded local damage, local damping loss, and an operational-only negative-control case. The results show that SPECTRA detects abrupt, gradual, nonlinear and damping-related damage mechanisms, while avoiding persistent false alarms under operational variability alone. Across the accepted benchmark suite, the persistent decision rule gives zero pre-damage persistent false alarms, finite detection delay in damage cases, and zero persistent alarms in the no-damage negative-control case. The framework provides a reproducible route for testing physics-informed anomaly inference before deployment in infrastructure digital twins. 
\end{abstract}


\section{Introduction}
\label{sec:introduction}

Structural health monitoring (SHM) aims to support engineering decisions by using measured structural response to infer whether a structure is behaving as expected or whether a structural change has occurred. Classical SHM methods have often been framed as pattern recognition problems, where features are extracted from vibration, strain or displacement data and then used to identify deviations from a reference condition \cite{worden2007machine,farrar2012shm}. This view has been highly influential because it provides a clear route from sensing to diagnosis. However, practical infrastructure monitoring requires more than statistical separation between healthy and abnormal feature patterns. The decision must also be physically meaningful, robust to normal variability and interpretable to engineers responsible for safety-critical assets \cite{brownjohn2007structural}. This requirement becomes more demanding in real-time monitoring, where the decision variable must be updated as new measurements arrive and where an isolated statistical change should not be treated, on its own, as evidence of structural damage. A persistent difficulty in SHM is that measured structural response changes for many reasons that are not damage. Temperature, humidity, traffic, wind, boundary condition changes, sensor noise and operational loading can all alter the measured response of a structure \cite{sohn2007eov}. The Z24 bridge study remains a widely cited example showing that environmental effects can produce modal changes that are comparable to, or larger than, changes caused by damage \cite{peeters2001z24}. Therefore, a monitoring method that reacts only to statistical novelty may confuse benign environmental and operational variability with structural deterioration. This is a fundamental limitation when SHM is intended to support infrastructure management, resilience planning and real-time decision-making.

The issue becomes more important as SHM moves towards digital twin-based infrastructure monitoring. A digital twin aims to provide a computational representation of the asset that can be compared with observations and updated as new data become available. Early digital twin concepts for structural life prediction were developed in aerospace applications \cite{tuegel2011digital}, and recent work has extended the discussion towards civil engineering systems, distributed sensing and model updating \cite{torzoni2024digital}. These developments create a strategic opportunity for SHM: monitoring decisions can be supported not only by statistical learning, but also by disagreement between measured behaviour and a physics-informed healthy model \cite{nath2026physics}.

This paper develops this idea through SPECTRA, a physics-informed eigen-compressed digital twin framework for real-time structural anomaly inference. \textbf{SPECTRA} stands for \textbf{S}elf-adaptive \textbf{P}hysics-informed \textbf{E}igen-\textbf{C}ompressed \textbf{T}win for \textbf{R}eal-time \textbf{A}nomaly inference. The framework is designed around a simple principle: a structural anomaly should not be inferred from statistical novelty alone; it interpreted together with physics-space disagreement between the measured response and a healthy twin. In this setting, the healthy twin provides the expected response of the structure under the assumed undamaged condition, while the measured response provides the actual behaviour of the monitored system. The difference between these two responses is treated as a twin innovation signal.

The proposed framework combines five main components. \textit{First}, a healthy physics-informed structural twin is used as the reference model. \textit{Second}, the system response is represented using an eigen-compressed dynamic basis, so that the dominant structural behaviour is retained in a compact form. \textit{Third}, residual and spectral features are extracted from measured and twin-based response quantities. \textit{Fourth}, kernel principal component analysis (KPCA) is used to form nonlinear monitoring indices. KPCA is suitable here because it extends principal component analysis to nonlinear feature spaces through a kernel eigenvalue problem \cite{scholkopf1998kpca}. \textit{Fifth}, a persistent decision rule is applied so that isolated excursions do not directly become structural alarms.

SPECTRA is motivated by the fact that real structures may exhibit different types of change. Some changes are abrupt, such as sudden stiffness loss. Others are gradual, such as progressive degradation. Crack-like behaviour can be nonlinear and non-smooth, for example through bilinear breathing stiffness \cite{friswell2002crack}. Damage may also occur under environmental and operational variability, where the structural change must be separated from benign global effects. In addition, some damage mechanisms may mainly affect damping rather than stiffness. These cases are challenging because they do not produce a single universal signature. Nonlinear structural dynamics and nonlinear system identification have long shown that such behaviours require careful interpretation of response features and model disagreement \cite{kerschen2006nonlinear}. The main contribution of this paper is therefore not the use of KPCA alone, nor the use of a digital twin alone. The contribution is the integration of a healthy physics-informed twin, eigen-compressed dynamics, twin-innovation features, KPCA monitoring indices and a persistent decision rule into a single anomaly inference framework. This integration gives the method a clear decision structure: statistical novelty is useful, but the final interpretation is strengthened when it is consistent with physics-space disagreement from the healthy twin.

The framework is assessed through seven numerical benchmarks. The benchmark suite includes a two-degree-of-freedom Duffing-type nonlinear drift case, sudden local stiffness loss, gradual stiffness degradation, bilinear breathing stiffness, environmental and operational variability-confounded local damage, local damping loss and an operational-only negative-control case. The negative-control case is included deliberately. It tests whether the framework avoids persistent structural alarms when the operating condition changes but no structural damage is introduced. This is necessary because a practical SHM method must demonstrate both sensitivity to damage and specificity under benign variability. The results show that SPECTRA detects abrupt, gradual, nonlinear and damping-related damage mechanisms with finite detection delay, while suppressing persistent false alarms in no-damage cases. The damage cases are not interpreted using one monitoring channel only. Instead, the results are analysed by considering the relationship between \(T^2\), SPE, the innovation index and the final persistent alarm. This provides a direct and technically interpretable route from measured response to anomaly decision.

The remainder of the paper is organised as follows. Section~\ref{sec:spectra_framework} presents the
SPECTRA framework, including the healthy twin, eigen-compressed representation, feature construction,
KPCA monitoring indices and the persistent decision rule. Section~\ref{sec:numerical_studies} defines
the seven benchmark cases and presents the numerical results case by case.
Section~\ref{sec:discussion} discusses the role of the different monitoring channels, the effect of
environmental and operational variability, the interpretation of the negative-control case, and the
main limitations. Section~\ref{sec:conclusions} summarises the findings and outlines future work.

\section{SPECTRA framework}
\label{sec:spectra_framework}

\begin{figure}[H] 
	\centering  \includegraphics[width=\textwidth]{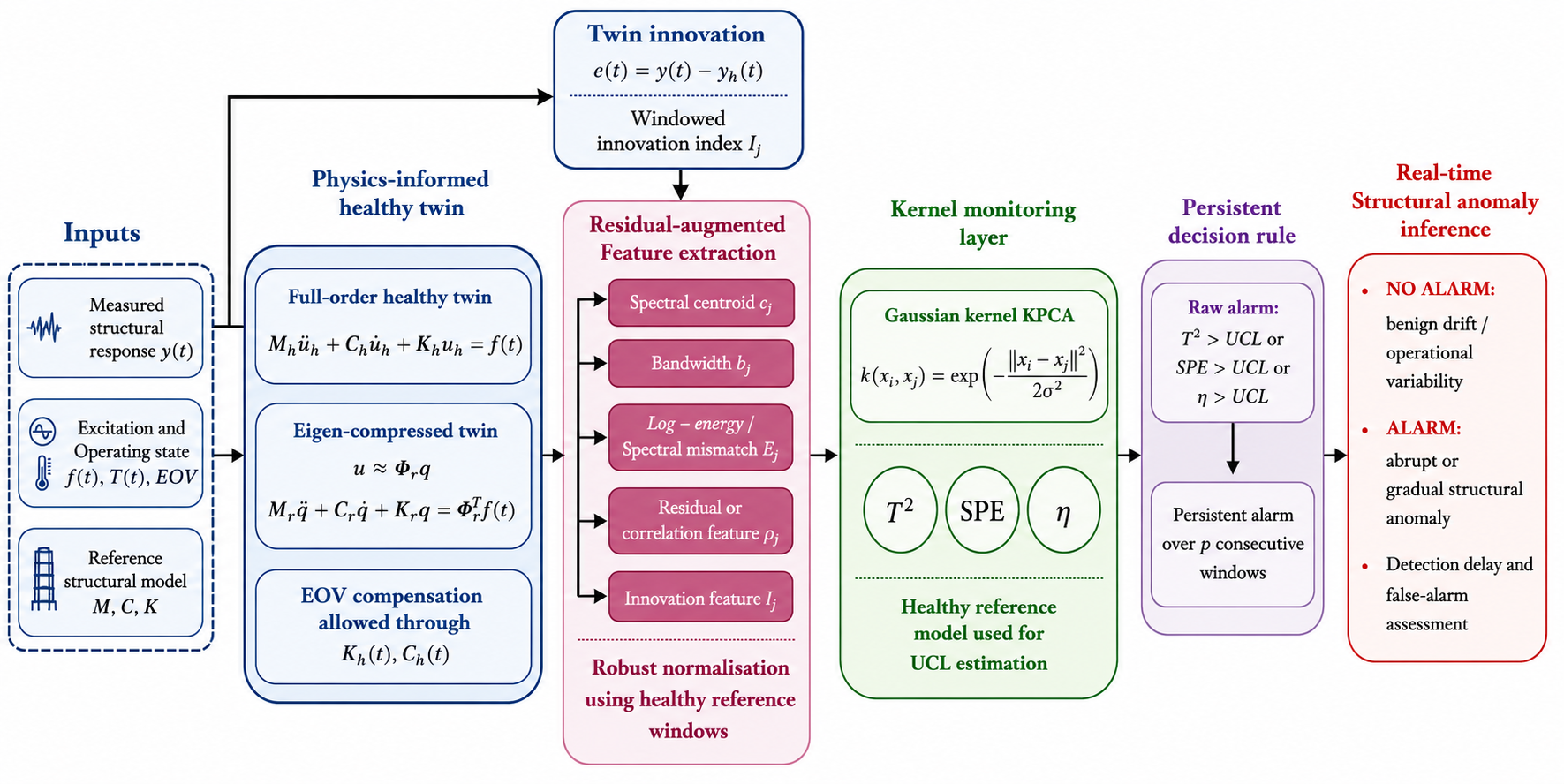}
	\caption{Schematic of the SPECTRA framework. The measured structural response is compared with a physics-informed healthy twin. Eigen-compressed dynamics, spectral features, twin innovation and KPCA monitoring indices are combined using a persistent decision rule.} 
	\label{fig:spectra_framework} 
\end{figure}

SPECTRA is designed as a physics-informed anomaly inference procedure in which the measured structural response is compared with the response predicted by a healthy structural twin. The framework follows the broader direction of digital twin-based civil infrastructure monitoring, where computational models, sensing data and decision-support processes are combined to support structural assessment and management \cite{torzoni2024digital}. SPECTRA does not rely on statistical novelty alone. Instead, it combines model-based disagreement, eigen-compressed dynamic representation, residual-augmented spectral features, nonlinear monitoring indices and a persistent decision rule. The complete workflow is shown in Fig.~\ref{fig:spectra_framework}.

\subsection{Full-order structural dynamics}
\label{subsec:full_order_dynamics}

The monitored structure is represented by the full-order dynamic equilibrium equation
\begin{equation}
	\mathbf{M}\ddot{\mathbf{u}}(t)
	+ \mathbf{C}(t)\dot{\mathbf{u}}(t)
	+ \mathbf{K}(t)\mathbf{u}(t)
	+ \mathbf{g}\big(\mathbf{u}(t),\dot{\mathbf{u}}(t),t\big)
	=
	\mathbf{f}(t),
	\label{eq:full_order_dynamics}
\end{equation}
where \(\mathbf{M}\), \(\mathbf{C}(t)\) and \(\mathbf{K}(t)\) are the mass, damping and stiffness matrices, respectively. The displacement vector is denoted by \(\mathbf{u}(t)\), while \(\mathbf{f}(t)\) is the external excitation. This form follows the standard matrix representation of structural dynamics for multi-degree-of-freedom systems. The term \(\mathbf{g}\big(\mathbf{u}(t),\dot{\mathbf{u}}(t),t\big)\) represents possible nonlinear contributions, including stiffness nonlinearity, breathing-type behaviour or other non-smooth restoring forces.

In practical SHM applications, the measured response is available only at selected sensor locations. Let \(y(t)\) denote the measured scalar response used for monitoring, such as the acceleration at a selected degree of freedom \cite{ostachowicz2019optimization}. SPECTRA treats this response as the observed behaviour of the real structure and compares it with the corresponding response predicted by a healthy twin.

\subsection{Physics-informed healthy twin}
\label{subsec:healthy_twin}

The healthy twin is a physics-informed reference model representing the expected undamaged behaviour of the structure. Its governing equation is written as
\begin{equation}
	\mathbf{M}\ddot{\mathbf{u}}_h(t)
	+ \mathbf{C}_h(t)\dot{\mathbf{u}}_h(t)
	+ \mathbf{K}_h(t)\mathbf{u}_h(t)
	=
	\mathbf{f}(t),
	\label{eq:healthy_twin}
\end{equation}
where \(\mathbf{u}_h(t)\) is the displacement response predicted by the healthy twin. The matrices \(\mathbf{C}_h(t)\) and \(\mathbf{K}_h(t)\) describe the damping and stiffness of the structure under the assumed healthy condition. The subscript \(h\) is used to indicate that the model corresponds to the healthy structural state.

For cases involving environmental and operational variability, the healthy stiffness matrix is allowed to vary with temperature through an environmental compensation factor. This is necessary because environmental and operational conditions can strongly affect measured structural response and may otherwise be confused with damage \cite{sohn2007eov}. The compensated healthy stiffness matrix is written as
\begin{equation}
	\mathbf{K}_h(t) = \mu_K(T(t))\mathbf{K}_{20},
	\label{eq:eov_stiffness}
\end{equation}
where \(\mathbf{K}_{20}\) is the healthy stiffness matrix at the reference temperature of \(20\,^{\circ}\mathrm{C}\), \(T(t)\) is the temperature at time \(t\), and \(\mu_K(T(t))\) is a scalar stiffness multiplier. This term allows benign global stiffness changes to be represented inside the healthy twin rather than being misinterpreted as damage.

The healthy twin is not used as a complete proof of structural condition. It is used as a physically meaningful baseline. If the measured structure remains consistent with the healthy twin, the innovation between measured and predicted response should remain small. If damage or another structural anomaly occurs, this innovation should increase because the healthy model no longer explains the measured response \cite{carden2004vibration}.

\subsection{Eigen-compressed dynamic representation}
\label{subsec:eigen_compressed_representation}

SPECTRA uses an eigen-compressed representation to retain the dominant dynamic behaviour in a compact form. For the healthy system, the undamped eigenvalue problem is defined as
\begin{equation}
	\mathbf{K}_h \boldsymbol{\Phi}
	=
	\mathbf{M}\boldsymbol{\Phi}\boldsymbol{\Omega}^2,
	\label{eq:eigenproblem}
\end{equation}
where \(\boldsymbol{\Phi}\) contains the mode shapes and \(\boldsymbol{\Omega}\) contains the corresponding natural circular frequencies. This eigenproblem and the associated modal representation follow the standard formulation used in structural dynamics. A reduced basis \(\boldsymbol{\Phi}_r\) is then formed using the first \(r\) dominant modes.

The displacement response can be approximated as
\begin{equation}
	\mathbf{u}(t) \approx \boldsymbol{\Phi}_r \mathbf{q}(t),
	\label{eq:modal_projection}
\end{equation}
where \(\mathbf{q}(t)\) is the reduced modal coordinate vector. Substitution of Eq.~\eqref{eq:modal_projection} into the linear part of Eq.~\eqref{eq:full_order_dynamics} gives the reduced-order equation
\begin{equation}
	\mathbf{M}_r \ddot{\mathbf{q}}(t)
	+ \mathbf{C}_r \dot{\mathbf{q}}(t)
	+ \mathbf{K}_r \mathbf{q}(t)
	=
	\boldsymbol{\Phi}_r^{T}\mathbf{f}(t),
	\label{eq:reduced_dynamics}
\end{equation}
where
\[
\mathbf{M}_r = \boldsymbol{\Phi}_r^{T}\mathbf{M}\boldsymbol{\Phi}_r,
\quad
\mathbf{C}_r = \boldsymbol{\Phi}_r^{T}\mathbf{C}\boldsymbol{\Phi}_r,
\quad
\mathbf{K}_r = \boldsymbol{\Phi}_r^{T}\mathbf{K}\boldsymbol{\Phi}_r.
\]
This eigen-compressed representation supports efficient real-time inference because the monitoring features are derived from a compact dynamic description rather than from the full structural state.

\subsection{Twin innovation}
\label{subsec:twin_innovation}

The central model-based quantity in SPECTRA is the twin innovation. Let \(y(t)\) denote the measured response and \(y_h(t)\) denote the corresponding response predicted by the healthy twin. The innovation is defined as
\begin{equation}
	e(t) = y(t) - y_h(t).
	\label{eq:innovation}
\end{equation}
This signal measures the disagreement between the real structure and the healthy twin in the response space used for monitoring.

For a monitoring window \(\mathcal{W}_j\), the innovation root-mean-square value is calculated as
\begin{equation}
	I_j =
	\sqrt{
		\frac{1}{|\mathcal{W}_j|}
		\sum_{t_i \in \mathcal{W}_j} e^2(t_i)
	}.
	\label{eq:innovation_rms}
\end{equation}
The index \(I_j\) provides a direct physics-space disagreement measure. A statistical feature may change under operational variability, but a persistent increase in \(I_j\) indicates that the measured response is no longer consistent with the healthy model. This model-based innovation is related in spirit to recursive eigen-perturbation residuals used for online damage detection \cite{bhowmik2019foep}. The distinction is that SPECTRA compares the measured response with a physics-informed healthy twin, rather than with a data-driven subspace estimated only from the measurements.

\subsection{Residual-augmented feature vector}
\label{subsec:feature_vector}

SPECTRA constructs one feature vector for each monitoring window. The feature vector combines spectral information, response-shape information and twin-innovation information. In this study, the generic feature vector is written as
\begin{equation}
	\mathbf{x}_j =
	\left[
	c_j,
	b_j,
	E_j,
	\rho_j,
	I_j
	\right],
	\label{eq:feature_vector}
\end{equation}
where \(c_j\) is the spectral centroid, \(b_j\) is a spectral bandwidth measure, \(E_j\) is a response energy measure, \(\rho_j\) is a residual or response-ratio feature, and \(I_j\) is the innovation index from Eq.~\eqref{eq:innovation_rms}. The exact feature composition can be adapted to the benchmark case, but the common principle is that each vector contains both statistical response descriptors and physics-informed innovation information. Adaptive multivariate decomposition methods can also supply such features for unsupervised
monitoring.

Before nonlinear monitoring, the features are robustly scaled using statistics from a healthy reference set:
\begin{equation}
	\tilde{x}_{j,k}
	=
	\frac{x_{j,k} - \mathrm{median}(x_{k}^{\mathrm{ref}})}
	{1.4826\,\mathrm{MAD}(x_{k}^{\mathrm{ref}}) + \epsilon}.
	\label{eq:robust_scaling}
\end{equation}
Here, \(x_{j,k}\) is the \(k\)-th feature in the \(j\)-th window, \(x_{k}^{\mathrm{ref}}\) denotes the corresponding feature values in the reference set, \(\mathrm{MAD}\) is the median absolute deviation, and \(\epsilon\) is a small positive constant used to avoid division by zero. The factor \(1.4826\) is the standard normal-consistency scaling used with MAD-type robust scale estimation \cite{rousseeuw1993mad}. Robust scaling is used because monitoring features may contain isolated excursions even under healthy operation.

\subsection{Kernel principal component monitoring}
\label{subsec:kpca_monitoring}

The scaled feature vectors are analysed using kernel principal component analysis. KPCA maps the input features into a nonlinear feature space and performs principal component analysis in that space using kernel evaluations \cite{scholkopf1998kpca}. In SPECTRA, the Gaussian radial basis function kernel is used:
\begin{equation}
	k(\mathbf{x}_i,\mathbf{x}_j)
	=
	\exp\left(
	-\frac{\|\mathbf{x}_i-\mathbf{x}_j\|^2}{2\sigma^2}
	\right),
	\label{eq:rbf_kernel}
\end{equation}
where \(\sigma\) is the kernel bandwidth.

Two KPCA monitoring indices are used. The first is a Hotelling-type statistic:
\begin{equation}
	T_j^2 =
	\sum_{\ell=1}^{r_k}
	\frac{z_{j,\ell}^2}{\lambda_\ell},
	\label{eq:t2}
\end{equation}
where \(z_{j,\ell}\) is the projection of the \(j\)-th feature vector onto the \(\ell\)-th retained kernel principal component, \(\lambda_\ell\) is the associated eigenvalue, and \(r_k\) is the number of retained kernel components. The \(T^2\) index measures variation within the retained nonlinear feature manifold and follows the broader principle of Hotelling-type multivariate monitoring \cite{vanhatalo2015effect}.

The second index is the squared prediction error:
\begin{equation}
	Q_j =
	\|\boldsymbol{\phi}(\mathbf{x}_j)
	-
	\hat{\boldsymbol{\phi}}(\mathbf{x}_j)\|^2.
	\label{eq:spe}
\end{equation}
Here, \(\boldsymbol{\phi}(\mathbf{x}_j)\) is the nonlinear feature-space representation and \(\hat{\boldsymbol{\phi}}(\mathbf{x}_j)\) is its reconstruction using the retained KPCA subspace. The \(Q_j\) index measures out-of-manifold novelty. This residual-type interpretation is consistent with the use of squared prediction error and residual control procedures in PCA-based monitoring \cite{shlens2014tutorial,krishnan2018rpca}.

The distinction between the monitoring indices is important. The Hotelling-type \(T^2\) index captures variations within the learned feature manifold. The squared prediction error (SPE) captures out-of-manifold novelty. The innovation index captures mismatch between the measured structure and the healthy physics-informed twin. These quantities provide complementary information. A large KPCA index may indicate feature novelty, but it does not automatically prove structural damage. A large innovation index indicates that the measured response is no longer consistent with the healthy twin. The persistent rule then checks whether this disagreement remains over consecutive windows. This structure is intended to reduce false alarms while preserving sensitivity to structural change \cite{hwang2009survey,chandola2009anomaly}.

\subsection{Normalised indices and persistent decision rule}
\label{subsec:persistent_decision_rule}

Upper control limits are estimated from the healthy reference windows. The three monitoring quantities are then normalised as
\begin{equation}
	\bar{T}_j^2 = \frac{T_j^2}{T^2_{\mathrm{UCL}}},
	\quad
	\bar{Q}_j = \frac{Q_j}{Q_{\mathrm{UCL}}},
	\quad
	\bar{\eta}_j = \frac{I_j}{I_{\mathrm{UCL}}},
	\label{eq:normalised_indices}
\end{equation}
where \(T^2_{\mathrm{UCL}}\), \(Q_{\mathrm{UCL}}\) and \(I_{\mathrm{UCL}}\) are the corresponding
upper control limits. A raw alarm requires the physics-space innovation to exceed its limit,
corroborated by nonlinear feature-space novelty:
\begin{equation}
	a_j =
	\mathbb{I}
	\left[
	\bar{\eta}_j > 1
	\;\wedge\;
	\left( \bar{T}_j^2 > 1 \;\lor\; \bar{Q}_j > 1 \right)
	\right],
	\label{eq:raw_alarm}
\end{equation}
where \(\mathbb{I}[\cdot]\) is the indicator function. Gating the alarm on the innovation channel is
deliberate: statistical novelty in \(\bar{T}_j^2\) or \(\bar{Q}_j\) can arise from benign operational
change, so it is treated as corroborating evidence rather than as sufficient grounds for an alarm.
An alarm is raised only when the measured response is inconsistent with the healthy twin and the
feature vector also departs from its healthy manifold.

A raw alarm is not immediately interpreted as a structural anomaly. SPECTRA applies a persistent decision rule:
\begin{equation}
	A_j =
	1
	\quad
	\mathrm{if}
	\quad
	\sum_{m=j-p+1}^{j} a_m = p.
	\label{eq:persistent_alarm}
\end{equation}
Here, \(p\) is the required number of consecutive alarmed windows. The final decision variable \(A_j\) is equal to one only when the raw alarm remains active for \(p\) consecutive monitoring windows. In operational-only negative-control cases, the SPE index may rise above its limit when the operating
regime changes, but the persistent damage flag remains zero because the innovation channel that gates
the alarm stays within its limit while the structure remains consistent with the healthy twin.

\subsection{Interpretation of anomaly evidence}
\label{subsec:anomaly_interpretation}

The final SPECTRA decision is based on the joint interpretation of three evidence channels. The first channel is the \(T^2\) index, which captures within-manifold variation. The second channel is the SPE index, which captures out-of-manifold novelty. The third channel is the innovation index, which captures disagreement between the measured response and the physics-informed healthy twin.

This distinction gives the framework a specific decision structure. A large \(T^2\) or SPE value may indicate that the feature distribution has changed, but this can occur under benign operational variability. A large innovation index indicates that the healthy twin no longer explains the measured response. A persistent alarm indicates that the disagreement is not isolated. Therefore, SPECTRA interprets a structural anomaly as a persistent and physically meaningful departure from the healthy twin, supported where appropriate by nonlinear feature-space novelty.

This structure is used consistently across the benchmark suite. In stiffness-loss cases, the SPE and innovation channels are expected to be informative. In gradual degradation cases, the innovation index should increase progressively after the onset of damage. In breathing-stiffness cases, the innovation signal should capture non-smooth response disagreement. In damping-loss cases, the innovation channel may dominate even when the KPCA indices remain below their limits. In operational-only negative-control cases, feature excursions may occur, but the persistent damage flag should remain zero if the structure remains consistent with the healthy twin.

\section{Benchmark suite and numerical studies}

\label{sec:numerical_studies}

The numerical studies evaluate the ability of the proposed framework to separate structural
anomalies from benign response changes. The benchmark suite contains seven controlled simulations:
a nonlinear two-degree-of-freedom response with an eigen-compressed twin representation, an abrupt
local stiffness reduction, a gradual local stiffness degradation, a bilinear breathing-stiffness
mechanism, a local stiffness loss under environmental and operational variability, a local
damping-loss mechanism, and an operational-only negative-control case. The purpose is not only to
demonstrate detection sensitivity, but also to show that persistent alarms are not produced by
every change in the measured response. Throughout, the healthy twin is given the reference
structural matrices and the same excitation as the monitored system, and a small deterministic
noise floor is added to the measured record only. This is deliberate: it isolates the decision
logic from estimation error in the twin, so that the benchmark tests the inference rule rather than
the quality of a fitted model. The consequences of relaxing this assumption are discussed in
Section~\ref{sec:discussion}.

Across the simulations, the monitored response is compared against a healthy reference twin. The
resulting innovation signal is combined with spectral features and kernel principal component
analysis indices. The final decision is based on persistent threshold exceedance rather than on an
isolated excursion. This is important because short excursions in feature space can arise from
excitation changes, spectral leakage, environmental variability or finite-window effects. A
physically meaningful alarm should therefore be supported by sustained incompatibility with the
healthy model. The benchmark suite is summarised in Tables~\ref{tab:numerical_cases} and~\ref{tab:model_parameters}. The first table defines the structural condition, disturbance and expected decision for each case, while the second table records the model and simulation parameters used to reproduce the numerical studies.

\begin{table}[t]
	\centering
	\caption{Summary of the numerical benchmark cases used to assess the proposed monitoring framework.}
	\label{tab:numerical_cases}
	\resizebox{\textwidth}{!}{%
	\begin{tabular}{llll}
		\hline
		Case & Structural condition & Main disturbance & Expected decision \\
		\hline
		Nonlinear response evolution & Healthy nonlinear system & Smooth response drift & No persistent alarm \\
		Abrupt stiffness loss & Damaged & 30\% stiffness loss at story 2 & Persistent alarm \\
		Gradual stiffness loss & Damaged & 30\% stiffness loss over time & Persistent alarm \\
		Breathing stiffness & Damaged & Bilinear stiffness at story 2 & Persistent alarm \\
		EOV-confounded stiffness loss & Damaged & Temperature variation and local damage & Persistent alarm \\
		Damping loss & Damaged & 82\% damping loss at story 2 & Persistent alarm \\
		Operational-only variability & Healthy & Temperature and force-regime change & No persistent alarm \\
		\hline
	\end{tabular}}
\end{table}

\begin{table}[t]
	\centering
	\caption{Model and simulation parameters for the seven benchmark cases. All shear-building
		models use a lumped-mass idealisation with story springs assembled into a tridiagonal stiffness
		matrix. Rayleigh damping is fitted to the target modal damping ratio $\zeta$.}
	\label{tab:model_parameters}
	\small
	\resizebox{\textwidth}{!}{%
	\begin{tabular}{lllll}
		\hline
		Case & DOF and mass & Stiffness & Damping & Sampling / duration \\
		\hline
		1 Duffing & 2-DOF, $m_1{=}1.00$, $m_2{=}0.80$~kg &
		$k_1{=}120$, $k_2{=}90$, $k_3{=}80$~N/m &
		viscous $c_1{=}0.80$, $c_2{=}0.35$, $c_3{=}0.60$ &
		$f_s{=}100$~Hz, $T{=}180$~s \\
		2 Step & 4-DOF, $1000\,[1,0.95,0.90,0.85]$~kg &
		$10^6[1.8,1.8,1.6,1.4]$~N/m & Rayleigh, $\zeta{=}0.02$ (modes 1,2) &
		$f_s{=}100$~Hz, $T{=}180$~s \\
		3 Gradual & 4-DOF, $1000$~kg per floor & $1.80\times10^6$~N/m per story &
		Rayleigh, $\zeta{=}0.02$ & $f_s{=}100$~Hz, $T{=}180$~s \\
		4 Breathing & 4-DOF, $1000$~kg per floor & $1.80\times10^6$~N/m per story &
		Rayleigh, $\zeta{=}0.02$ & $f_s{=}100$~Hz, $T{=}180$~s \\
		5 EOV & 4-DOF, $800$~kg per floor & $10^6[1.8,1.6,1.4,1.2]$~N/m &
		Rayleigh, $\zeta{=}0.02$ & $f_s{=}100$~Hz, $T{=}180$~s \\
		6 Damping & 4-DOF, $1000$~kg per floor & $1.80\times10^6$~N/m per story &
		local dampers $2.40$~kN\,s/m $+$ background $\zeta{=}0.003$ &
		$f_s{=}200$~Hz, $T{=}180$~s \\
		7 Negative control & operational-only, healthy & (as case 5 reference) &
		(as case 5 reference) & $f_s{=}100$~Hz, $T{=}180$~s \\
		\hline
	\end{tabular}}
\end{table}


\begin{figure}[H]
	\centering
	\fbox{\includegraphics[width=0.85\textwidth]{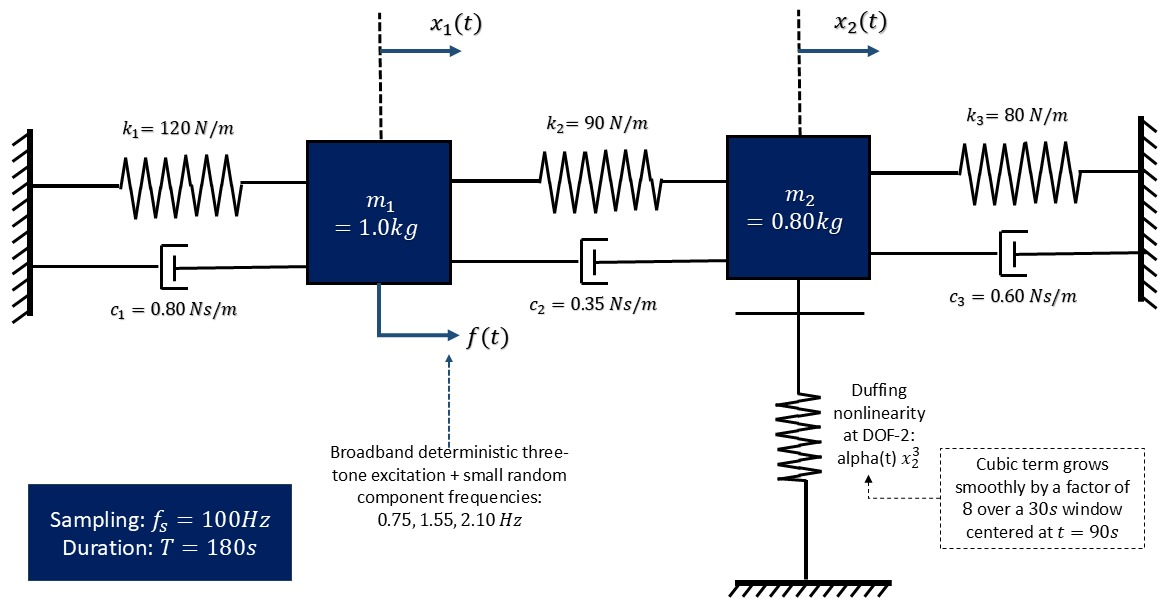}}
	\caption{Schematic of the nonlinear two-degree-of-freedom Duffing benchmark used in Case 1. The system consists of two lumped masses connected through linear spring-damper elements, with a cubic Duffing-type nonlinearity acting at DOF-2. The external force is applied at DOF-1 as a broadband deterministic three-tone excitation with a small random component. The cubic term grows smoothly over a 30~s window centred at \(t=90~\mathrm{s}\), so the case represents smooth nonlinear drift rather than a discrete damage event.}
	\label{fig:case1_duffing_schematic}
\end{figure}

\subsection{Nonlinear response evolution and eigen-compressed twin}
\label{subsec:nonlinear_eigen_compressed_twin}

The first case is a two-degree-of-freedom oscillator with a smooth cubic Duffing-type nonlinearity at the second degree of freedom, as shown in Fig.~\ref{fig:case1_duffing_schematic}. The linear stiffnesses are \(k_1=120~\mathrm{N/m}\), \(k_2=90~\mathrm{N/m}\) and \(k_3=80~\mathrm{N/m}\), with masses \(m_1=1.00~\mathrm{kg}\) and \(m_2=0.80~\mathrm{kg}\). The viscous damping coefficients are \(c_1=0.80~\mathrm{N\,s/m}\), \(c_2=0.35~\mathrm{N\,s/m}\) and \(c_3=0.60~\mathrm{N\,s/m}\). The cubic term grows smoothly by a factor of eight over a 30~s window centred on \(t=90~\mathrm{s}\), so the change is a gradual, amplitude-dependent stiffening rather than a discrete event. The excitation is a broadband deterministic three-tone force at \(0.75\), \(1.55\) and \(2.10~\mathrm{Hz}\), with a small random component, applied at the first degree of freedom and sampled at \(100~\mathrm{Hz}\) over \(180~\mathrm{s}\). The purpose of this case is not detection but specificity: it tests whether the framework stays quiet when the response drifts smoothly while the structure remains healthy, which is common in real assets that show amplitude-dependent behaviour.

\begin{figure}[H]
	\centering
	\begin{minipage}[t]{0.48\textwidth}
		\centering
		\includegraphics[width=\linewidth]{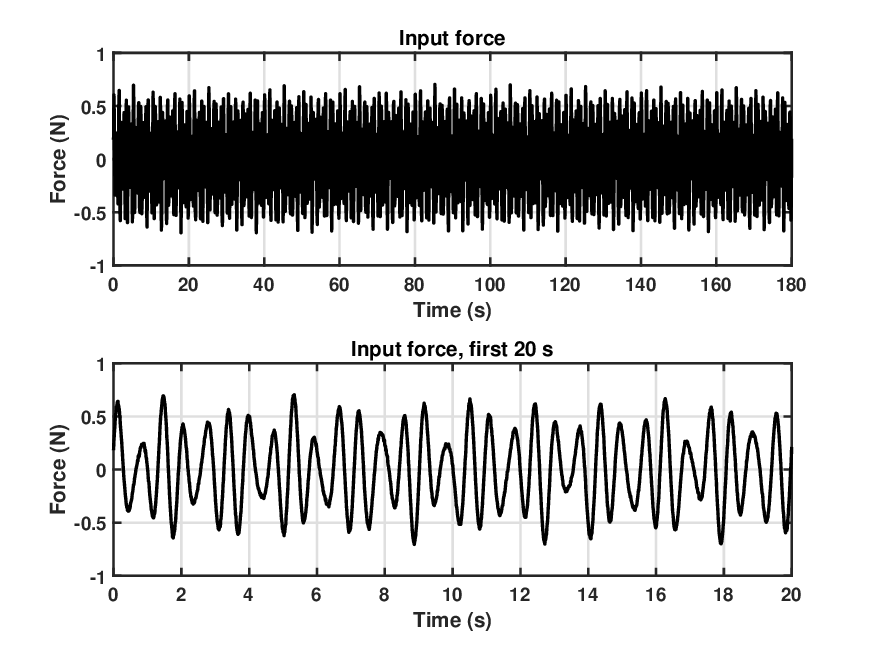}
		\caption{Input force used for the nonlinear two-degree-of-freedom simulation. The lower panel shows the first 20 s and illustrates the local time-scale of the applied excitation.}
		\label{fig:case1_input}
	\end{minipage}
	\hfill
	\begin{minipage}[t]{0.48\textwidth}
		\centering
		\includegraphics[width=\linewidth]{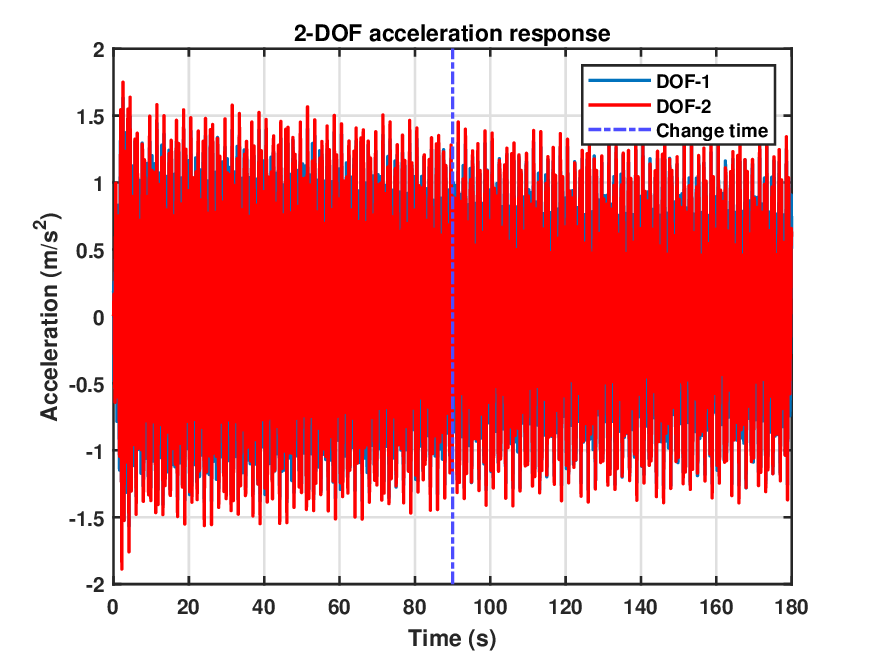}
		\caption{Acceleration response of the nonlinear two-degree-of-freedom system. The response evolves smoothly and does not contain a discrete structural damage event.}
		\label{fig:case1_acceleration}
	\end{minipage}
\end{figure}

Figures~\ref{fig:case1_input} and~\ref{fig:case1_acceleration} show the applied force and the two acceleration responses. A vertical marker denotes the nominal change time used only for feature tracking; no damage is introduced. The response amplitude and orbit evolve smoothly across this marker, which is the behaviour the monitoring layer must learn to tolerate.

\begin{figure}[H]
	\centering
	
	\begin{minipage}[t]{0.48\textwidth}
		\centering
		\includegraphics[width=\linewidth]{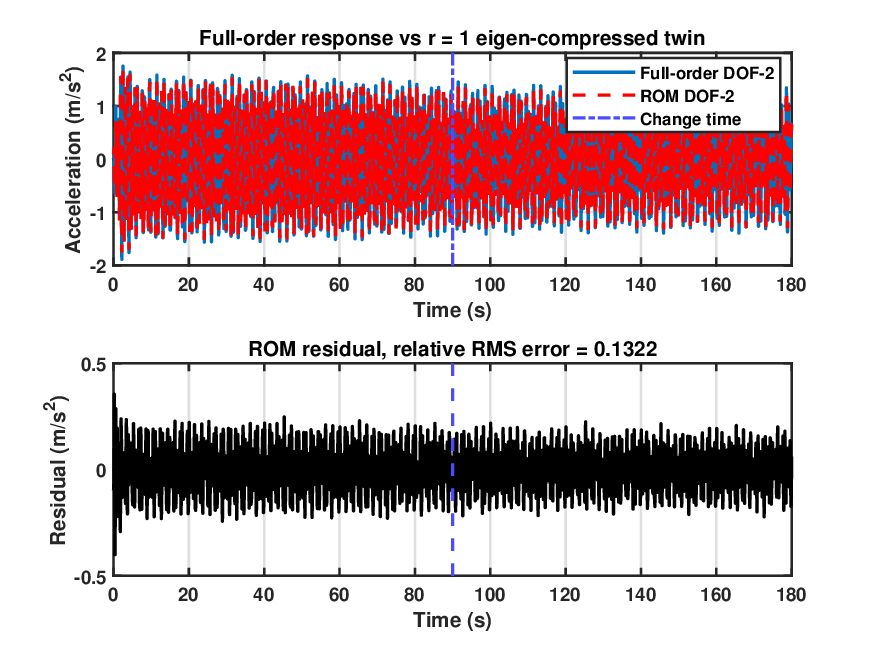}
		\caption{Full-order nonlinear response compared with the rank-one eigen-compressed twin. 
			The residual remains bounded and stationary, with a relative RMS error of $0.1322$, 
			which is suitable for monitoring-oriented analysis rather than exact response reconstruction.}
		\label{fig:case1_physics}
	\end{minipage}
	\hfill
	\begin{minipage}[t]{0.48\textwidth}
		\centering
		\includegraphics[width=\linewidth]{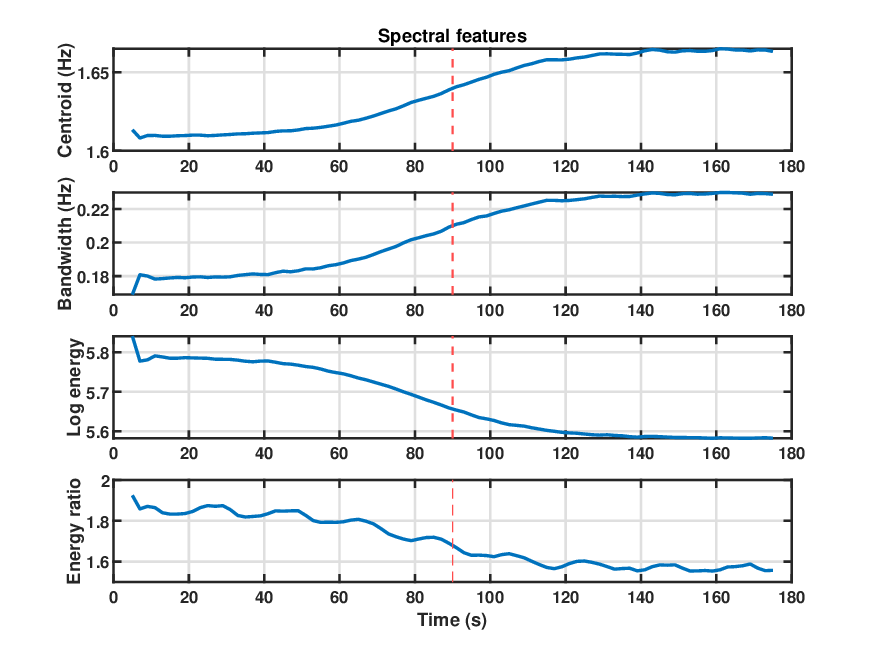}
		\caption{Spectral features extracted from the nonlinear response. 
			The features evolve smoothly and provide a controlled representation of non-damage-induced response drift.}
		\label{fig:case1_features}
	\end{minipage}
	
\end{figure}

Figure~\ref{fig:case1_physics} compares the full-order acceleration with the rank-one
eigen-compressed twin used inside SPECTRA. The residual is bounded and stationary, with a relative
root-mean-square error of $0.1322$. A residual of this size is expected when the response is
represented by a single dominant mode, and it is stationary across the nominal change marker, which
confirms that the smooth stiffening does not degrade the compression. This bounded, stationary
residual is precisely why the innovation channel alone is not used as a detector: the framework must
be able to separate this fixed compression error, which is present even in the healthy state, from
the growing model disagreement that accompanies genuine damage. 

\begin{figure}[H]
	\centering
	\begin{subfigure}[h]{0.48\textwidth}
		\centering
		\includegraphics[width=\linewidth]{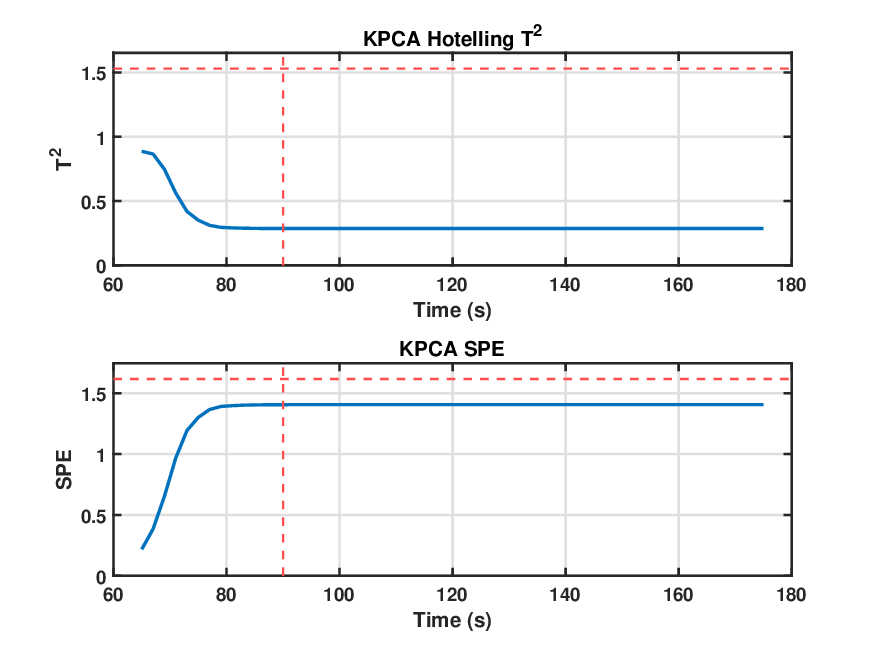}
		\caption{KPCA $T^2$ and SPE.}
		\label{fig:case1_kpca}
	\end{subfigure}
	\hfill
	\begin{subfigure}[h]{0.48\textwidth}
		\centering
		\includegraphics[width=\linewidth]{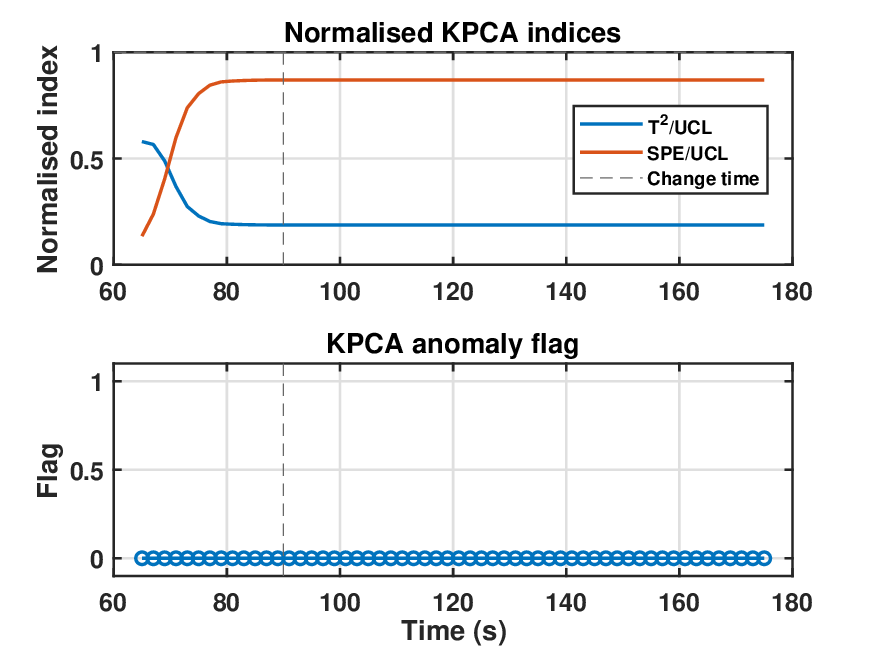}
		\caption{Normalised indices and flag.}
		\label{fig:case1_flags}
	\end{subfigure}
	\caption{Nonlinear response case, monitoring and decision. (a) Both KPCA indices remain below
		their upper control limits. (b) The normalised indices and the persistent flag stay at zero
		throughout, confirming that smooth nonlinear drift does not generate an alarm.}
	\label{fig:case1_decision}
\end{figure}

The extracted spectral features are shown in Figure~\ref{fig:case1_features}: the centroid and
bandwidth rise gradually while the energy features fall, tracking the stiffening nonlinearity. On
their own these trends look like a monotonic change, which is exactly the trap this case sets. The
KPCA indices and the persistent flag in Figure~\ref{fig:case1_decision} show the correct response:
the indices stay below their upper control limits and the flag remains at zero for the entire
record. While panel~\ref{fig:case1_kpca} confirms that the nonlinear feature trajectory remains inside the healthy KPCA control region, panel~\ref{fig:case1_flags} shows that the normalised indices do not generate a persistent alarm despite the smooth drift in the response. The framework therefore treats smooth nonlinear drift as healthy, because the twin
innovation stays bounded and the KPCA novelty never persists above threshold. This is the baseline
against which the damage cases are read.

\subsection{Abrupt local stiffness reduction}
\label{subsec:abrupt_stiffness_reduction}

The second case is a four-degree-of-freedom shear-building model with floor masses of
$1000\,[1.00,0.95,0.90,0.85]$~kg and story stiffnesses $10^6[1.80,1.80,1.60,1.40]$~N/m, damped by a
Rayleigh model fitted to $\zeta=0.02$ on the first two modes (Table~\ref{tab:model_parameters}). A
broadband force is applied at the first floor and the system is integrated at $100$~Hz over $180$~s
by a fourth-order Runge-Kutta scheme. Damage is a discrete $30\%$ loss of the second-story stiffness
at $t=90$~s, which is the clearest possible test of sensitivity to an abrupt structural change. A
deterministic noise floor at $0.3\%$ of the DOF-2 signal standard deviation is added to the measured
record.
\begin{figure}[H]
	\centering
	\begin{subfigure}[t]{0.48\textwidth}
		\centering
		\includegraphics[width=0.9\linewidth]{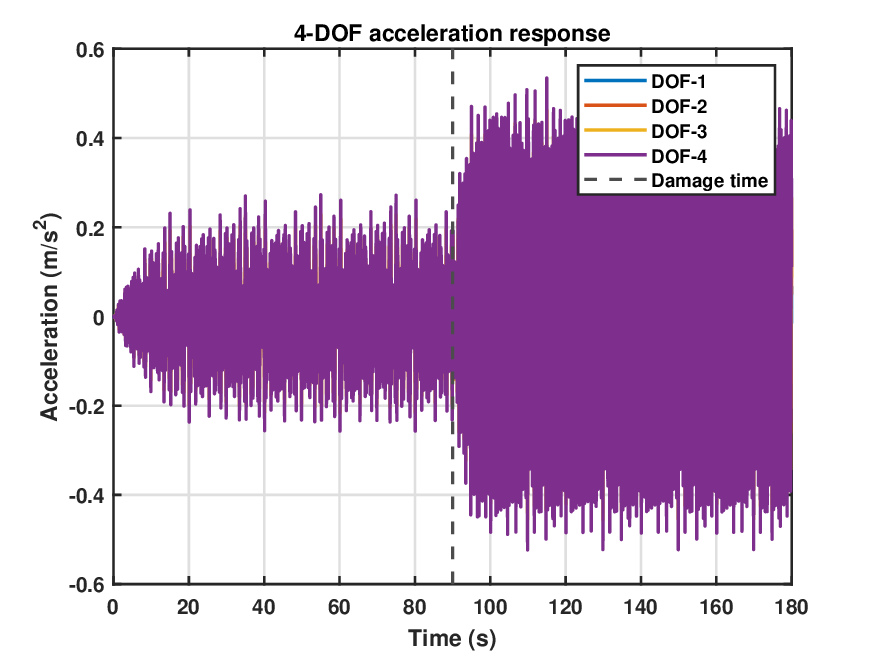}
		\caption{Acceleration response.}
		\label{fig:case2_acceleration}
	\end{subfigure}
	\hfill
	\begin{subfigure}[t]{0.48\textwidth}
		\centering
		\includegraphics[width=0.9\linewidth]{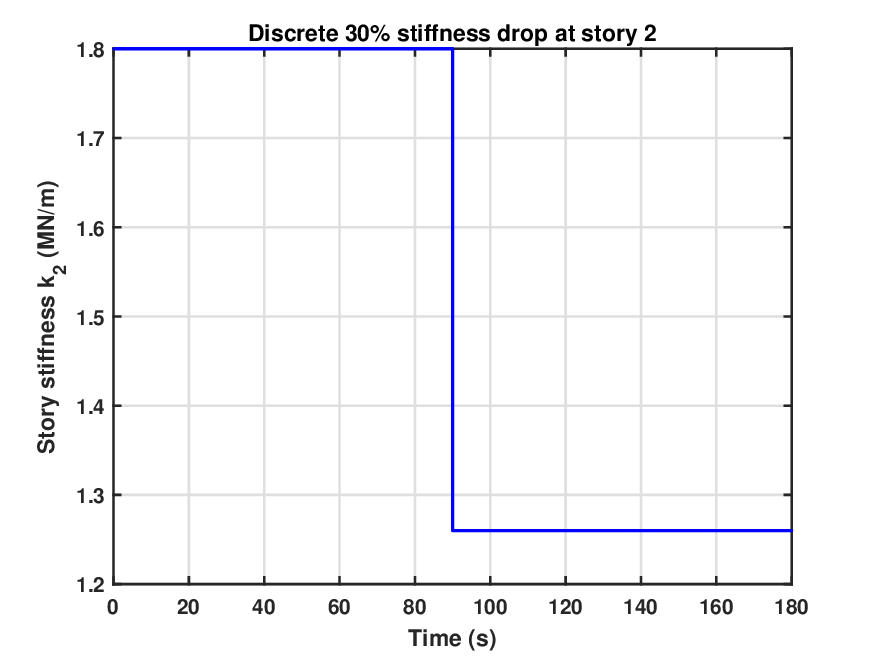}
		\caption{Imposed step in $k_2$.}
		\label{fig:case2_stiffness}
	\end{subfigure}
	\caption{Abrupt stiffness-loss case, setup. (a) The four-floor acceleration response, with the
		damage time marked; the amplitude rises after $t=90$~s. (b) The discrete $30\%$ reduction of the
		second-story stiffness.}
	\label{fig:case2_setup}
\end{figure}

\begin{figure}[H]
	\centering
	\includegraphics[width=0.55\textwidth]{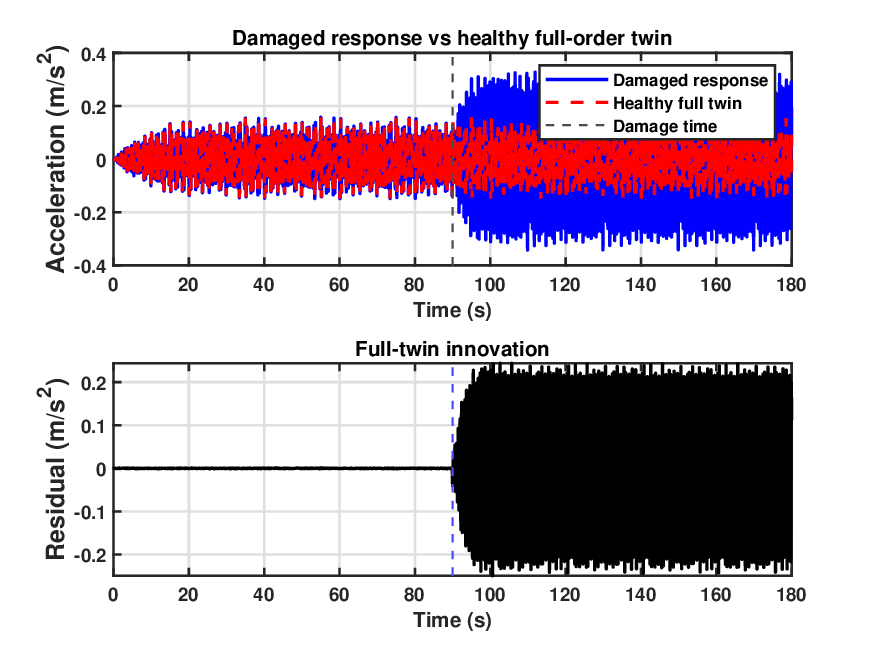}
	\caption{Damaged response compared with the healthy full-order twin for the abrupt stiffness-loss case. The full-twin innovation increases sharply after the local stiffness reduction.}
	\label{fig:case2_residual}
\end{figure}

Figure~\ref{fig:case2_acceleration} shows the
four-floor acceleration response, with the damage time marked. The response amplitude jumps at
$t=90$~s, most visibly in the upper floors, and Figure~\ref{fig:case2_stiffness} shows the imposed
step reduction in $k_2$. Figure~\ref{fig:case2_residual} compares the measured response with the healthy full-order twin.
Before $t=90$~s the innovation sits at the noise floor, with an RMS of $4.6442\times10^{-4}$; after
the stiffness drop it rises to $1.5959\times10^{-1}$, a change of roughly three orders of magnitude.
The measured structure is no longer compatible with the healthy twin. This separation is large
because, in these simulations, the twin uses the exact healthy matrices and the same excitation, so
the pre-damage residual reflects only the added noise. The magnitude of the jump should therefore
be read as confirmation that the decision logic fires on genuine model disagreement, not as an
estimate of the signal-to-noise ratio available in the field.

\begin{figure}[H]
	\centering
	\begin{subfigure}[t]{0.48\textwidth}
		\centering
		\includegraphics[width=\linewidth]{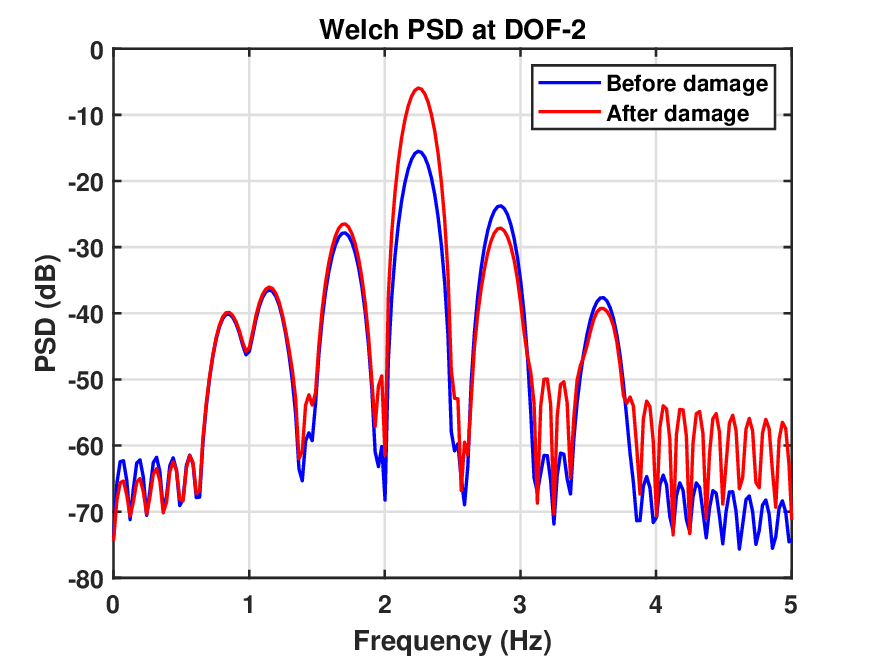}
		\caption{Welch PSD, before and after damage.}
		\label{fig:case2_psd}
	\end{subfigure}
	\hfill
	\begin{subfigure}[t]{0.48\textwidth}
		\centering
		\includegraphics[width=\linewidth]{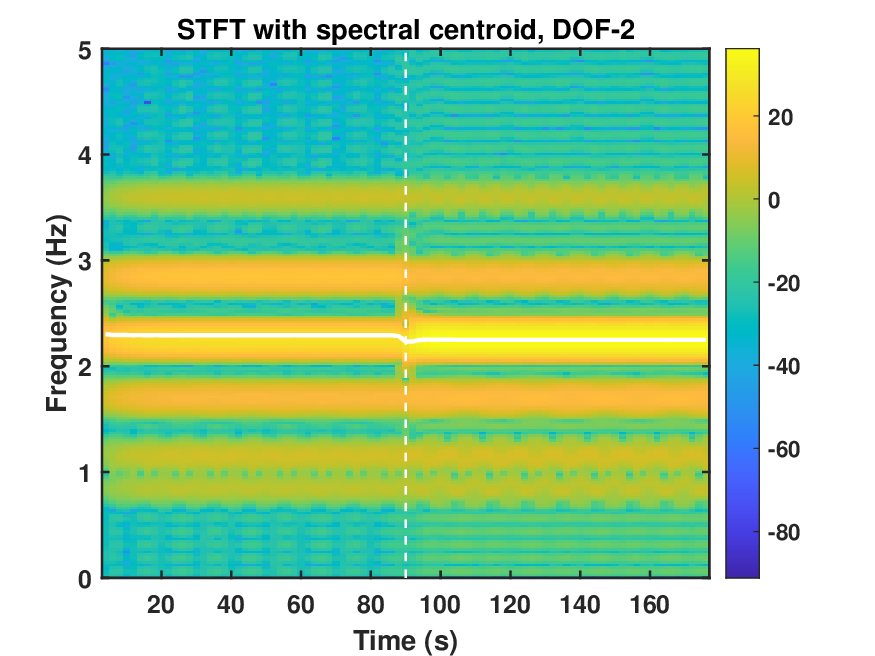}
		\caption{STFT with spectral centroid.}
		\label{fig:case2_stft}
	\end{subfigure}
	\caption{Abrupt stiffness-loss case, frequency-domain evidence at DOF-2. (a) The Welch spectrum
		shifts and redistributes energy after damage, with the dominant peak moving down in frequency.
		(b) The short-time Fourier transform shows this as an abrupt step in the spectral centroid at
		$t=90$~s, in contrast to the smooth drift of case~1.}
	\label{fig:case2_physics}
\end{figure}

The physics of the abrupt loss is seen most clearly in the frequency and time-frequency domains,
collected in Figure~\ref{fig:case2_physics}. The Welch spectrum (panel \ref{fig:case2_psd}) shows a clear
redistribution of response energy after the stiffness drop, with the dominant peak moving down in
frequency, exactly as expected when a story softens. The short-time Fourier transform (panel \ref{fig:case2_stft})
localises the same event in time, showing an abrupt step in the spectral centroid at the damage
instant rather than the smooth drift seen in the healthy nonlinear case. Read together, the two
panels place the event in both the frequency and time-frequency domains, which gives the statistical
decision that follows a clear physical basis. The state-space picture tells the same story: the
DOF-2 orbit expands after damage as the softened story admits larger relative displacement.

\begin{figure}[H]
	\centering
	\includegraphics[width=0.55\textwidth]{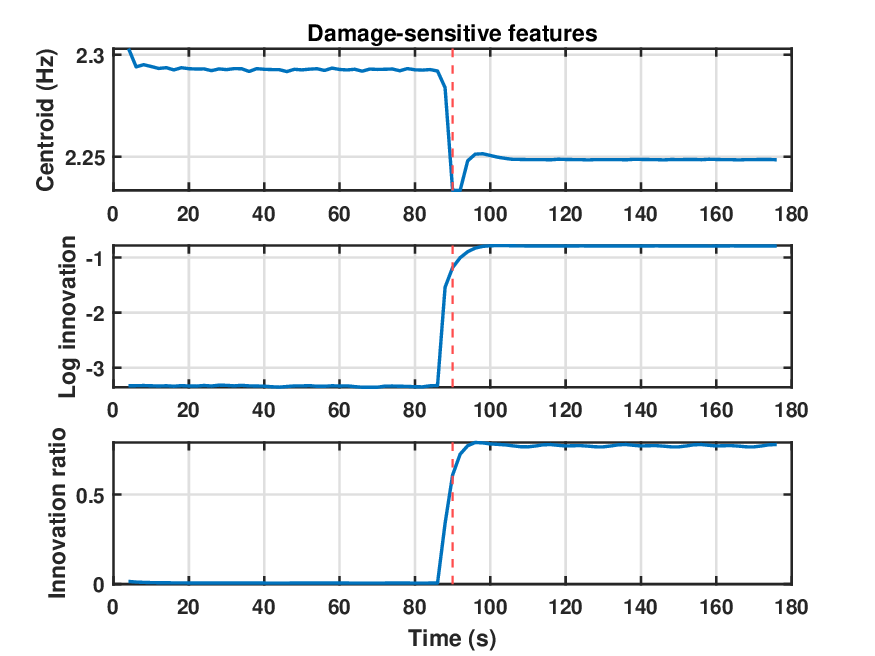}
	\caption{Abrupt stiffness-loss case: damage-sensitive features. The centroid and innovation
		features step sharply at the damage time.}
	\label{fig:case2_features}
\end{figure}

\begin{figure}[H]
	\centering
	\begin{subfigure}[h]{0.48\textwidth}
		\centering
		\includegraphics[width=\linewidth]{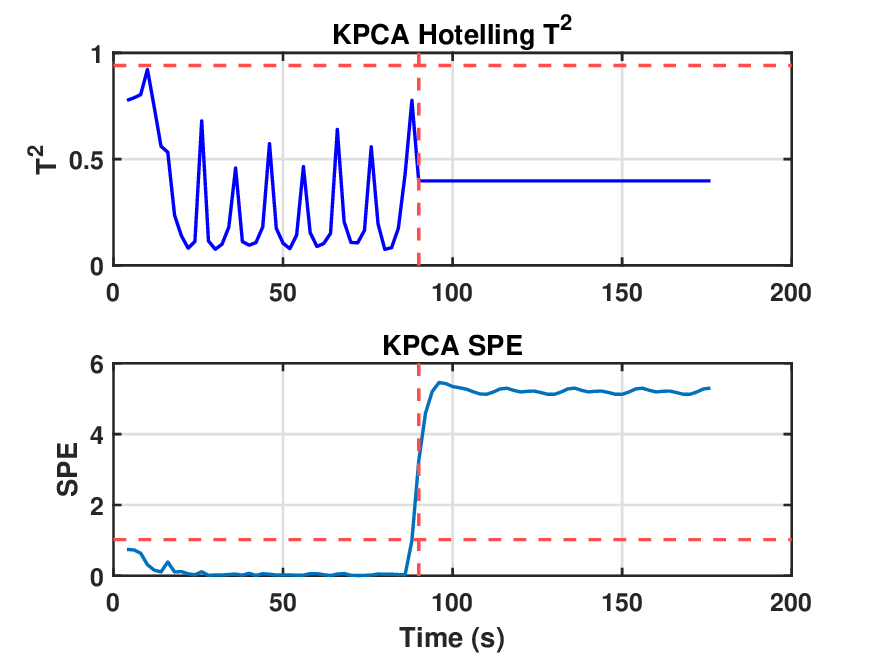}
		\caption{KPCA $T^2$ and SPE.}
		\label{fig:case2_kpca}
	\end{subfigure}
	\hfill
	\begin{subfigure}[h]{0.48\textwidth}
		\centering
		\includegraphics[width=\linewidth]{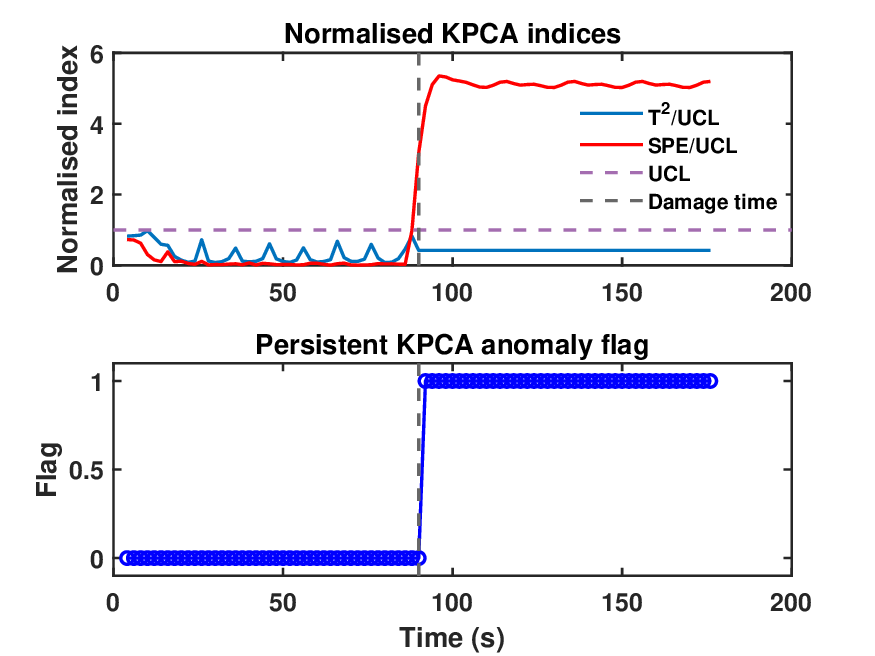}
		\caption{Normalised indices and flag.}
		\label{fig:case2_flags}
	\end{subfigure}
	\caption{Abrupt stiffness-loss case, monitoring and decision. (a) The SPE index exceeds its upper
		control limit and remains there, while $T^2$ is less responsive. (b) The normalised indices and
		the persistent flag show a sustained alarm after damage and no persistent pre-damage false alarm.}
	\label{fig:case2_decision}
\end{figure}

The feature-level response is shown in Figure~\ref{fig:case2_features}. The spectral centroid and the
innovation features step sharply at $t=90$~s, tracking the same event seen in the physics plots. The
step is largest in the innovation channel, because the healthy twin, which retains the undamaged
stiffness, can no longer reproduce the measured response once the story has softened. The monitoring indices and the final decision are shown in Figure~\ref{fig:case2_decision}. In the
KPCA indices (panel \ref{fig:case2_kpca}) the SPE crosses its upper control limit and stays above it, while the $T^2$
index is comparatively insensitive, because the post-damage feature vector leaves the healthy
manifold rather than moving within it. The normalised indices and the persistent flag (panel \ref{fig:case2_flags}) show
a sustained alarm from shortly after the damage time, with no persistent exceedance before it. The
finite detection delay is set by the monitoring window length and the two-window persistence
requirement, not by any tuning of thresholds.

\subsection{Gradual local stiffness degradation}
\label{subsec:gradual_stiffness_degradation}

The third case uses the same four-degree-of-freedom shear building as case~2, but with uniform
floor masses of $1000$~kg and uniform story stiffnesses of $1.80\times10^6$~N/m
(Table~\ref{tab:model_parameters}). The excitation is applied at the second floor. Damage is a
$30\%$ loss of the second-story stiffness introduced not as a step but as a linear ramp between
$t=90$~s and $t=130$~s, after which the reduced stiffness is held constant. This is a harder test than case~2 since the change develops over several monitoring windows, the features and the innovation must reveal a trend rather than a discontinuity. Progressive degradation is a natural setting for online damage detection methods, where the decision variable should respond to accumulating structural change rather than only to abrupt jumps \cite{sharma2024fire}. In SPECTRA, this requirement is tested through the twin innovation, which should grow in step with the imposed stiffness loss.

\begin{figure}[H]
	\centering
	\begin{subfigure}[h]{0.48\textwidth}
		\centering
		\includegraphics[width=\linewidth]{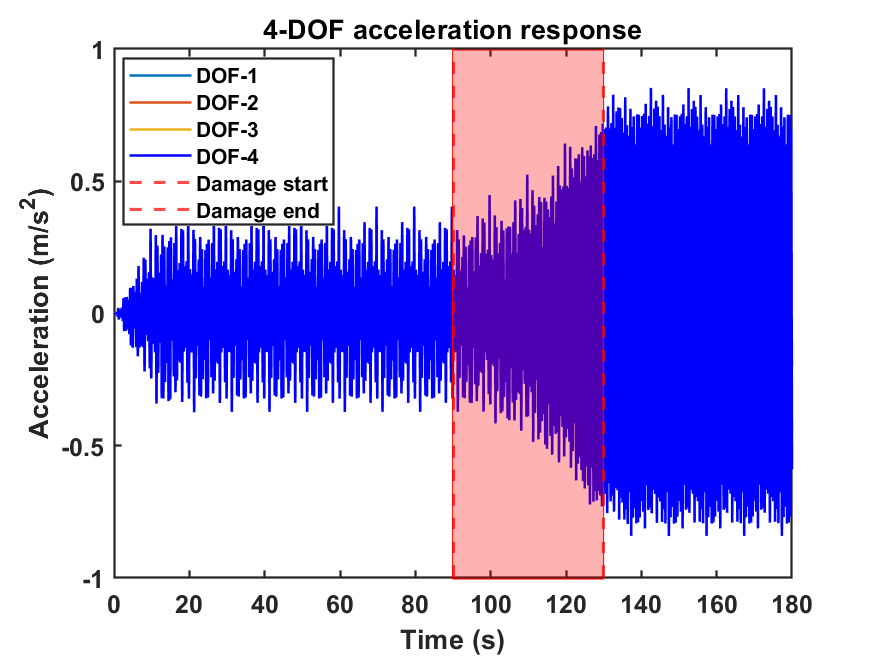}
		\caption{Acceleration response.}
		\label{fig:case3_acceleration}
	\end{subfigure}
	\hfill
	\begin{subfigure}[h]{0.48\textwidth}
		\centering
		\includegraphics[width=\linewidth]{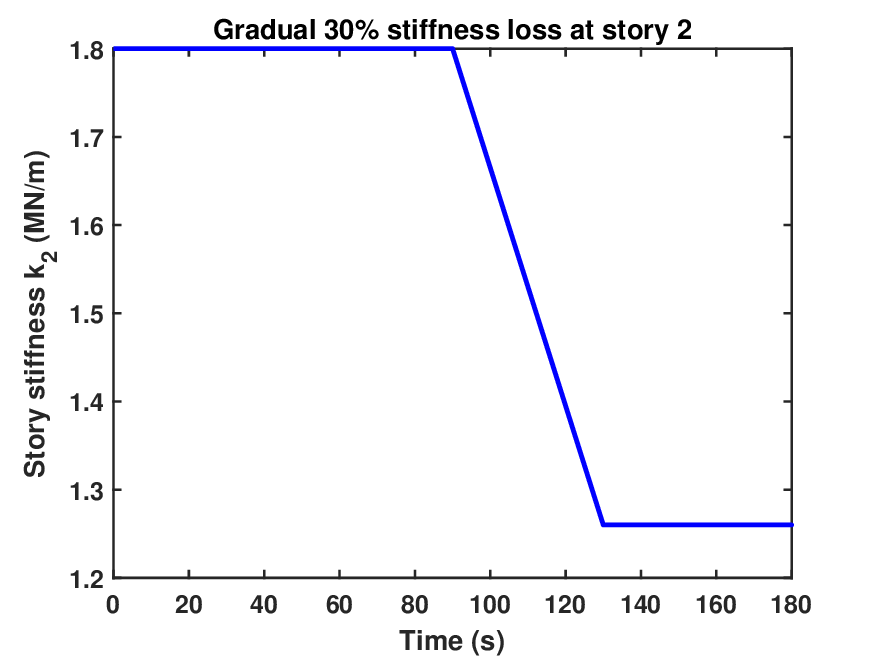}
		\caption{Ramped reduction in $k_2$.}
		\label{fig:case3_stiffness}
	\end{subfigure}
	\caption{Gradual stiffness-degradation case, setup. (a) The four-floor acceleration response
		grows through the degradation interval and then stabilises. (b) The second-story stiffness is
		reduced by $30\%$ as a linear ramp between $t=90$~s and $t=130$~s.}
	\label{fig:case3_setup}
\end{figure}

Figure~\ref{fig:case3_acceleration} shows the system response and
Figure~\ref{fig:case3_stiffness} shows the ramped reduction in $k_2$. The response amplitude grows
through the degradation interval and then settles at the new, larger level.


\begin{figure}[H]
	\centering
	
	\begin{minipage}[t]{0.48\textwidth}
		\centering
		\includegraphics[width=0.89\linewidth]{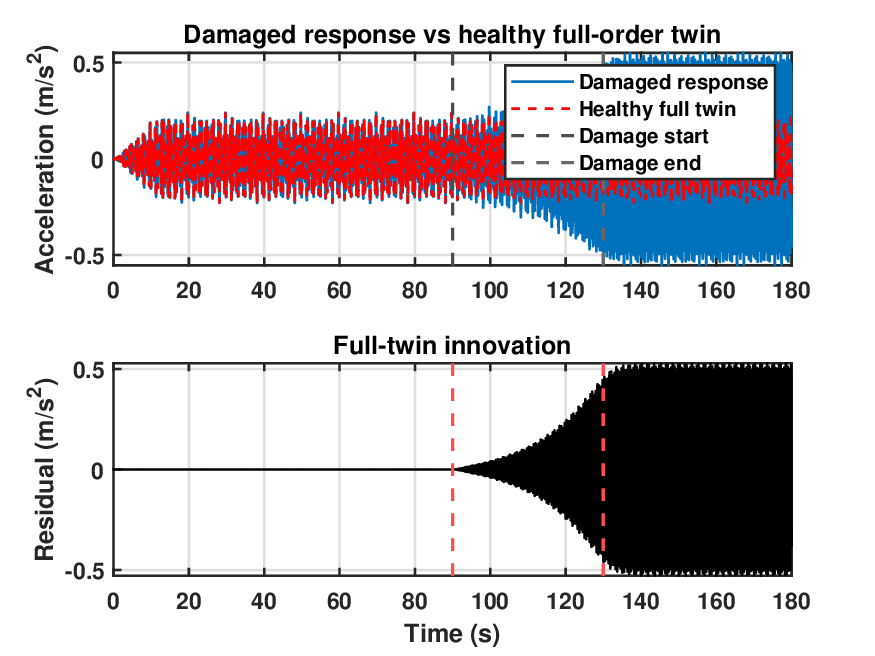}
		\caption{Damaged response compared with the healthy full-order twin for the gradual stiffness-degradation case. 
			The innovation increases progressively as the damage evolves.}
		\label{fig:case3_residual}
	\end{minipage}
	\hfill
	\begin{minipage}[t]{0.48\textwidth}
		\centering
		\includegraphics[width=0.88\linewidth]{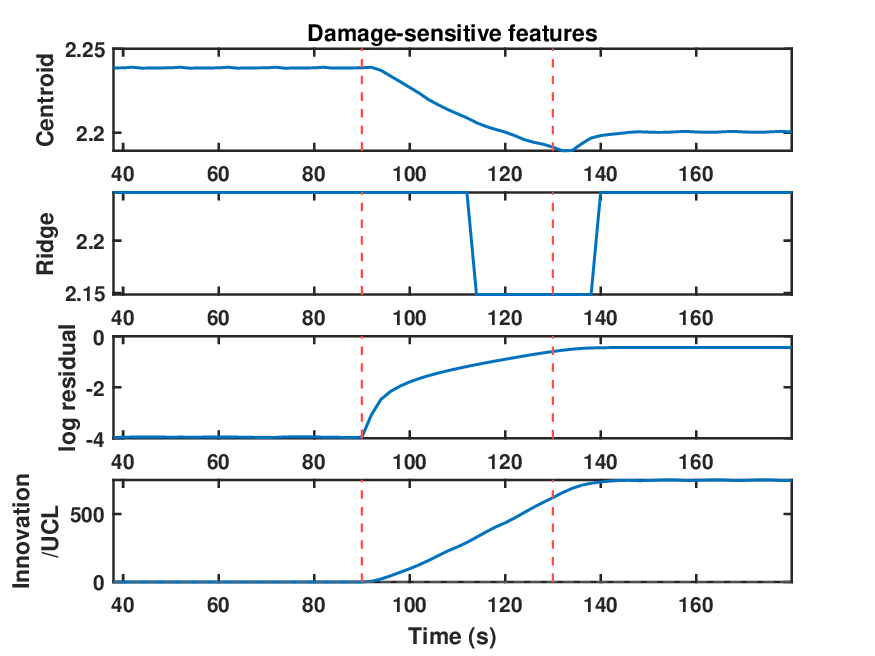}
		\caption{Damage-sensitive features for the gradual stiffness-degradation case. 
			The innovation feature rises progressively during the degradation interval and subsequently stabilises.}
		\label{fig:case3_features}
	\end{minipage}
	
\end{figure}

Figure~\ref{fig:case3_residual} compares the measured response with the healthy full-order twin. The
innovation grows progressively rather than stepping, from $1.1068\times10^{-4}$ before degradation
to $3.5965\times10^{-1}$ once it is complete. This shows that the twin tracks the accumulating
change: the residual accumulates in step with the stiffness loss rather than appearing all at once.
As in case~2, the absolute scale of this growth reflects the exact-twin setting; the point of
interest is the monotonic trend, which is what the decision layer must catch early rather than only
at the end. The physics of the gradual case mirrors the abrupt case but develops over the degradation interval
rather than at a single instant. The DOF-2 orbit expands progressively as the story softens, the
Welch spectrum gains low-frequency energy consistent with a more flexible structure, and the
time-frequency content migrates gradually during the marked interval rather than stepping. The
contrast with the abrupt case is the key point: the same monitoring layer must respond to a slow
trend here and to a discontinuity there.

\begin{figure}[H]
	\centering
	\begin{subfigure}[t]{0.48\textwidth}
		\centering
		\includegraphics[width=0.89\linewidth]{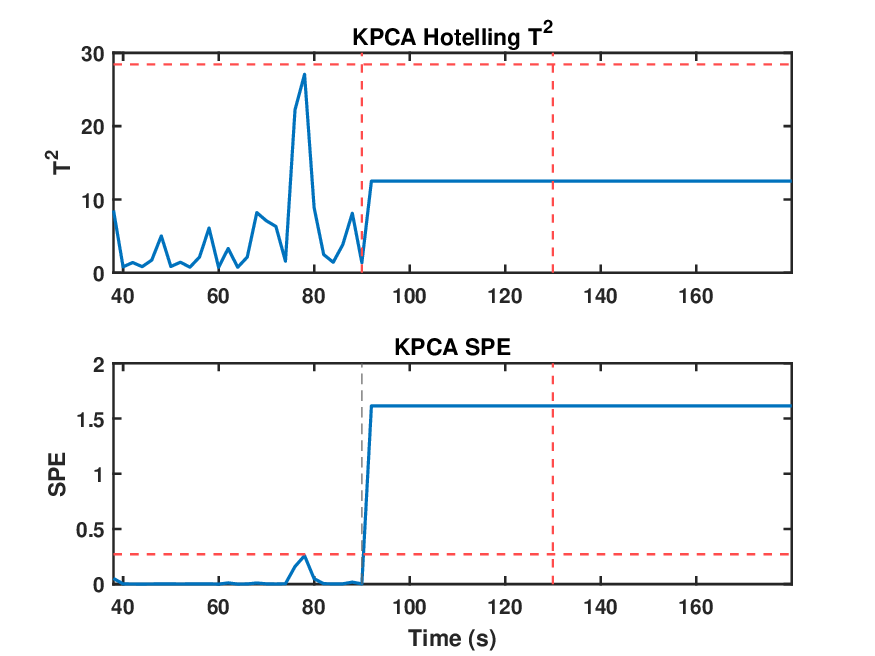}
		\caption{KPCA $T^2$ and SPE.}
		\label{fig:case3_kpca}
	\end{subfigure}
	\hfill
	\begin{subfigure}[t]{0.48\textwidth}
		\centering
		\includegraphics[width=0.89\linewidth]{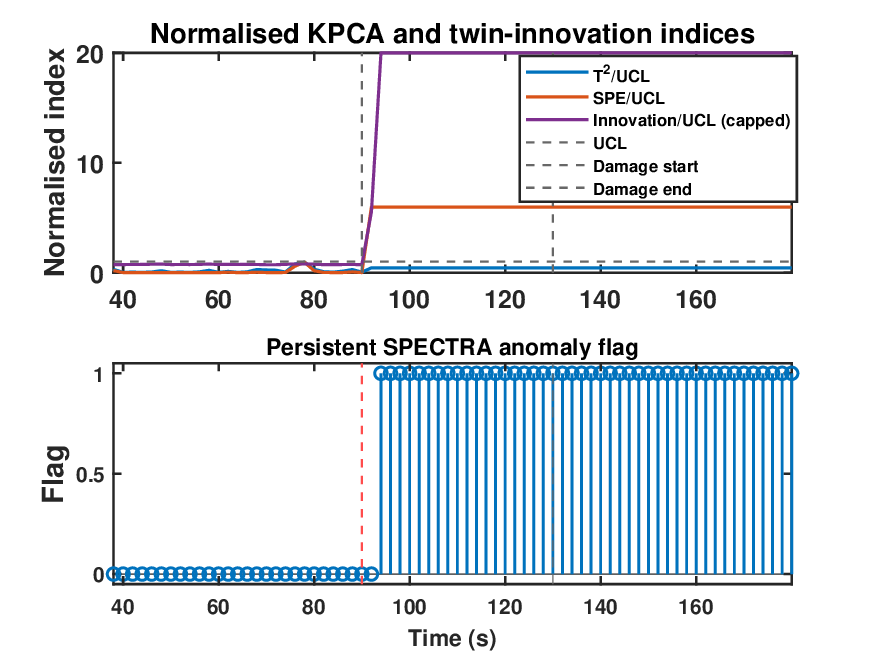}
		\caption{Normalised indices and flag.}
		\label{fig:case3_flags}
	\end{subfigure}
	\caption{Gradual stiffness-degradation case, monitoring and decision. (a) The SPE index and
		innovation channel separate from the healthy state as the damage accumulates. (b) The persistent
		flag activates after the onset of degradation and remains active, with no persistent false alarm
		beforehand.}
	\label{fig:case3_decision}
\end{figure}

The feature-level response is shown in Figure~\ref{fig:case3_features}. The innovation index rises
steadily from the onset of degradation and then plateaus, mirroring the imposed stiffness ramp,
while the spectral features drift over the same interval. The gradual rise, rather than a step,
is the signature the decision layer must catch.
The monitoring indices and the final decision are shown in Figure~\ref{fig:case3_decision}. The SPE
index and the innovation channel separate progressively from the healthy state as the damage
accumulates (panel \ref{fig:case3_kpca}), and the persistent flag activates shortly after onset and stays on, with no
persistent exceedance before $t=90$~s (panel \ref{fig:case3_flags}). The gradual nature of the change means that the detection delay is governed by how far the trend must
develop before the gated indices cross their limits for two consecutive windows. This is the expected trade-off between early warning and false-alarm control,
and it is the quantity a field deployment would tune.

\subsection{Bilinear breathing-stiffness behaviour}
\label{subsec:breathing_stiffness}

\begin{figure}[h]
	\begin{subfigure}[h]{0.48\textwidth}
		\centering
		\includegraphics[width=0.89\linewidth]{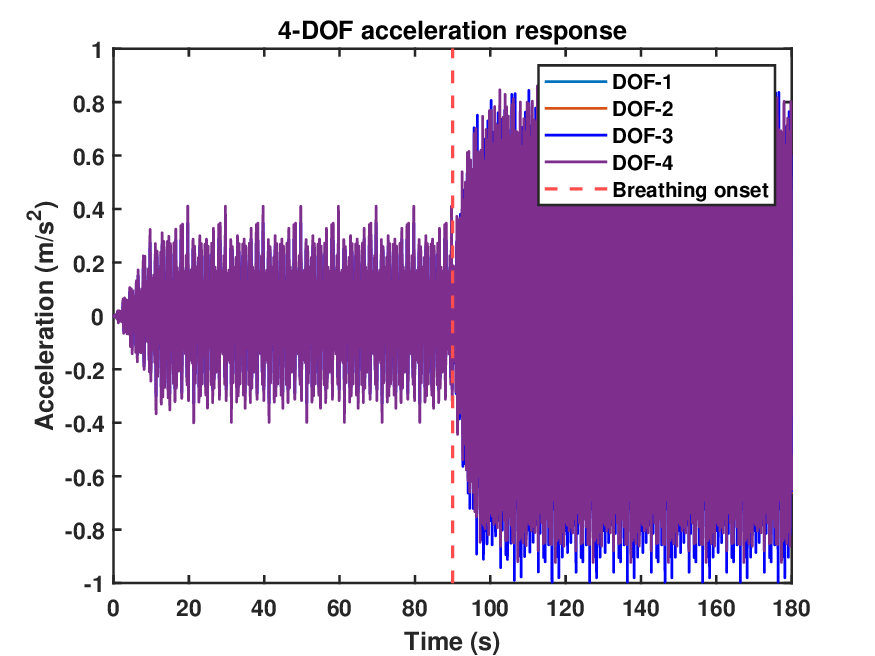}
		\caption{Acceleration response.}
		\label{fig:case4_acceleration}
	\end{subfigure}
	\hfill
	\begin{subfigure}[h]{0.48\textwidth}
		\centering
		\includegraphics[width=0.89\linewidth]{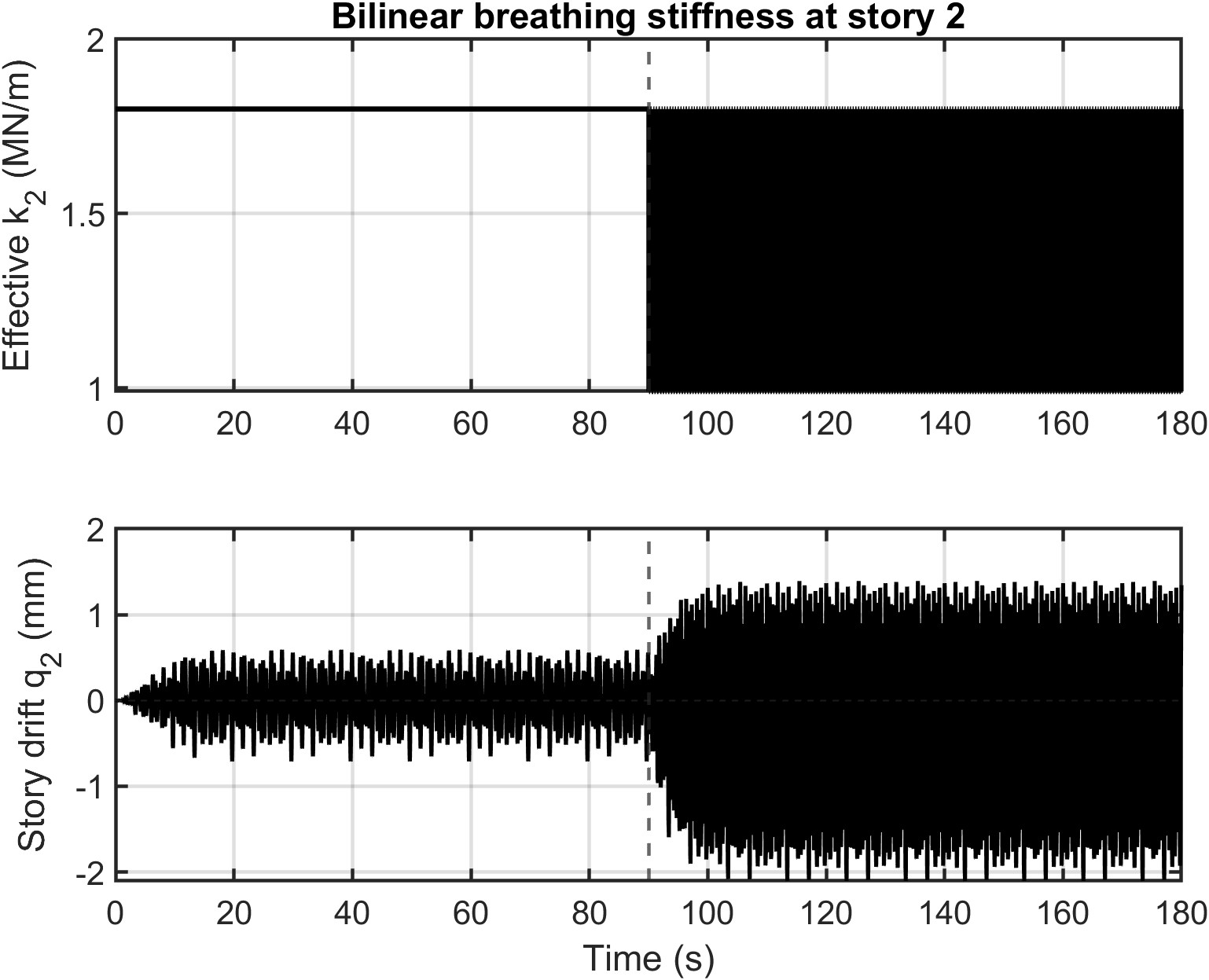}
		\caption{Effective $k_2$ and story drift.}
		\label{fig:case4_bilinear}
	\end{subfigure}
	\caption{Bilinear breathing-stiffness case, setup. (a) The broadband force used to activate the
		state-dependent response. (b) The four-floor acceleration response, with the breathing onset
		marked at $t=90$~s. (c) The effective second-story stiffness alternates between its closed and
		open values after onset, and the story drift increases accordingly.}
	\label{fig:case4_setup}
\end{figure}

The fourth case introduces a state-dependent, non-smooth nonlinearity at the second story of the
four-degree-of-freedom shear building. The masses are uniform at $1000$~kg per floor and the
undamaged story stiffnesses are uniform at $1.80\times10^{6}$~N/m, with Rayleigh damping fitted to
$\zeta=0.02$ on the first two modes (see Table~\ref{tab:model_parameters}). The excitation is a
broadband deterministic force applied at the second floor and the system is integrated at $100$~Hz
over $180$~s. From $t=90$~s the second-story spring becomes bilinear: its stiffness switches
between a closed value $k_{+}=1.80\times10^{6}$~N/m and an open value
$k_{-}=(1-0.45)k_{+}=0.99\times10^{6}$~N/m, so that the effective stiffness alternates within each
vibration cycle. This represents a breathing crack that opens and closes under load. It is the most
demanding case for a linear healthy twin, because the damaged response is not merely a shifted
version of the healthy response: it is generated by a mechanism the linear twin cannot represent at
all.

\begin{figure}[h]
	\centering
	\begin{subfigure}[t]{0.48\textwidth}
		\centering
		\includegraphics[width=0.9\linewidth]{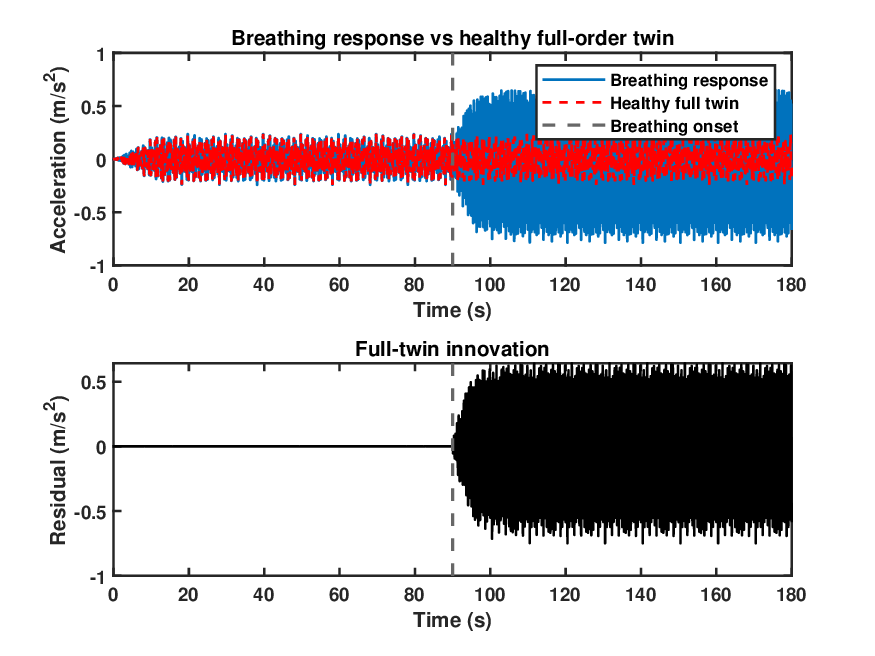}
		\caption{Twin innovation.}
		\label{fig:case4_residual}
	\end{subfigure}
	\hfill
	\begin{subfigure}[t]{0.48\textwidth}
		\centering
		\includegraphics[width=0.9\linewidth]{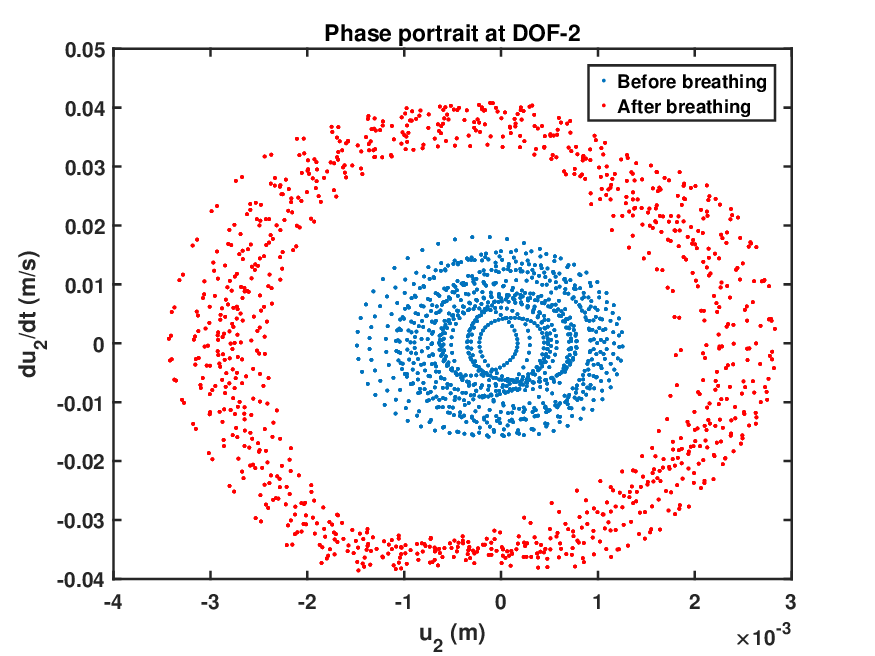}
		\caption{Phase portrait, DOF-2.}
		\label{fig:case4_phase}
	\end{subfigure}
	\caption{Bilinear breathing-stiffness case, physics evidence at DOF-2. (a) The healthy linear
		twin cannot reproduce the state-dependent response, so the innovation rises sharply after the
		breathing onset. (b) The phase portrait expands and loses its smooth single-orbit shape,
		consistent with a non-smooth restoring force.}
	\label{fig:case4_physics}
\end{figure}

The acceleration response and imposed bilinear stiffness are shown in
Figure~\ref{fig:case4_setup}. The effective story-2 stiffness switches state after the breathing
onset, and the associated story drift grows, which is the physical fingerprint of a crack that no
longer transmits load continuously across the cycle. The comparison with the healthy full-order twin is shown in Figure~\ref{fig:case4_physics}(a). The
innovation grows from $1.9546\times10^{-4}$ before breathing to $3.5709\times10^{-1}$ afterwards.
The twin is linear and time-invariant, so it cannot reproduce the state-dependent switching of the
breathing crack, and the residual is therefore not a small mismatch but a structurally different signal. The
phase portrait in Figure~\ref{fig:case4_physics}(b) confirms this in state space: after the onset the
orbit expands and loses its smooth single-loop shape, developing the corner-like features that are
characteristic of a non-smooth restoring force. The Welch spectrum, described here rather than
plotted, broadens correspondingly, spreading energy away from the original modal peaks. This
broadband spreading is why the innovation and SPE channels, not a single centroid shift, carry the
detection in this case.

\begin{figure}[H]
	\centering
	\begin{subfigure}[t]{0.48\textwidth}
		\centering
		\includegraphics[width=0.9\linewidth]{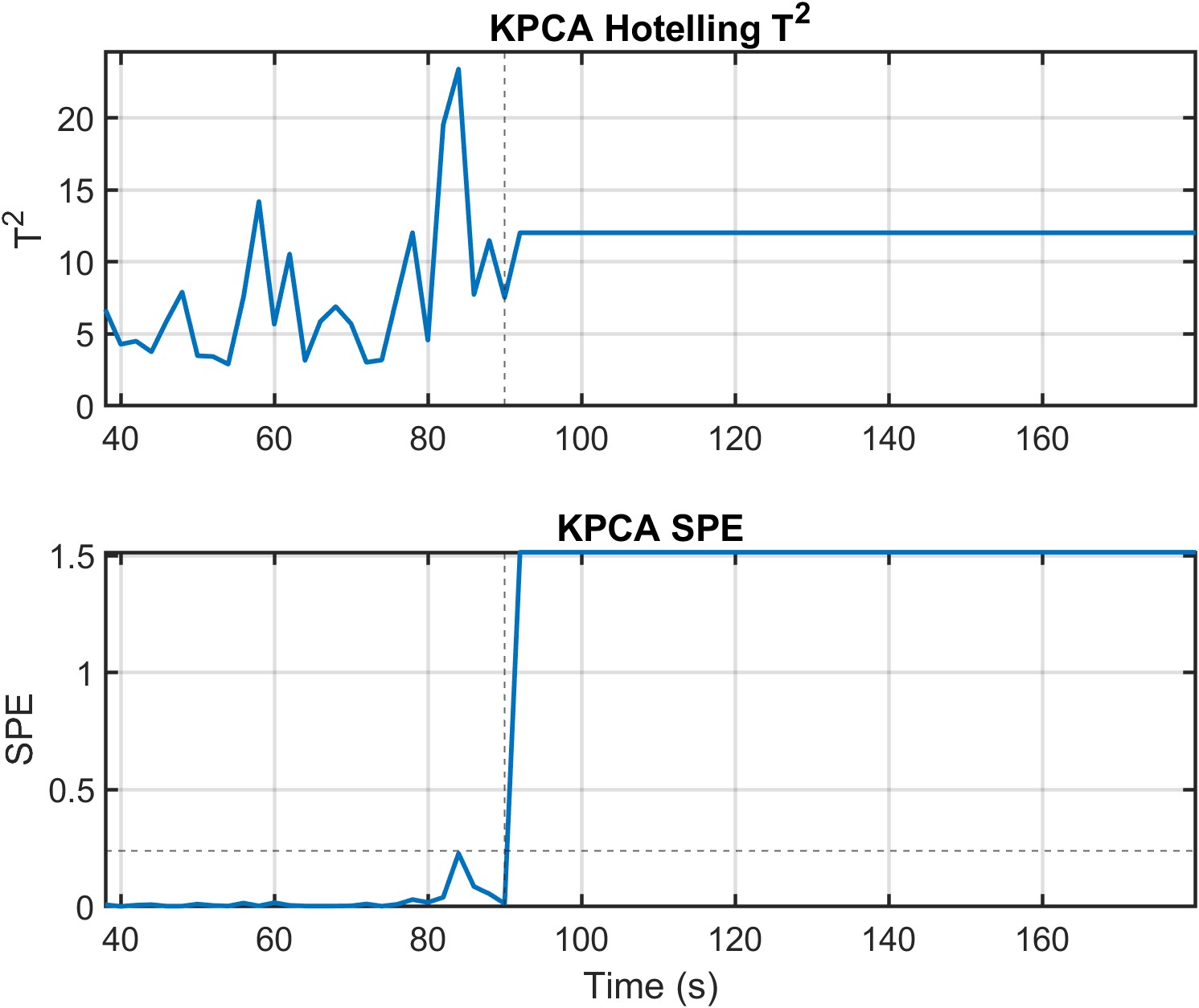}
		\caption{KPCA $T^2$ and SPE.}
		\label{fig:case4_kpca}
	\end{subfigure}
	\hfill
	\begin{subfigure}[t]{0.48\textwidth}
		\centering
		\includegraphics[width=0.9\linewidth]{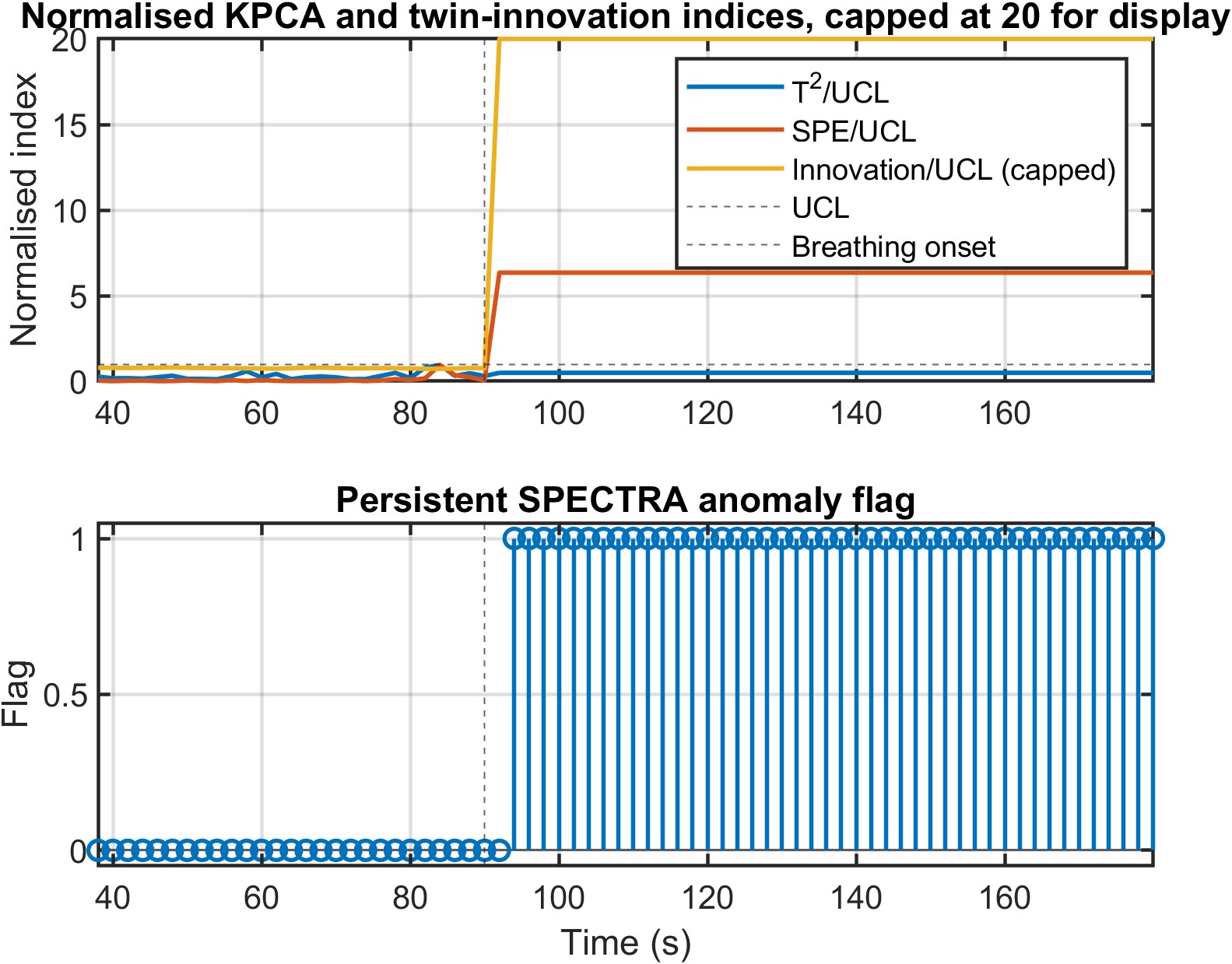}
		\caption{Normalised indices and flag.}
		\label{fig:case4_flags}
	\end{subfigure}
	\caption{Bilinear breathing-stiffness case, monitoring and decision. (a) The SPE index crosses
		its upper control limit while $T^2$ remains less sensitive. (b) The persistent flag produces a
		sustained alarm after onset with no pre-onset false alarm.}
	\label{fig:case4_decision}
\end{figure}
The feature-level and decision-level response is shown in Figure~\ref{fig:case4_decision}. The
residual-augmented feature vector for this case retains the innovation channels explicitly, because
breathing damage is fundamentally a model-mismatch phenomenon even when the ordinary spectral shift
is modest. The KPCA SPE index and the innovation index both exceed their upper control limits after
onset, while the $T^2$ index responds more weakly. The persistent flag activates shortly after the
breathing onset and remains active, with no persistent exceedance beforehand. This case therefore
demonstrates that the same decision logic that handled the linear stiffness-loss cases also
captures a nonlinear, non-smooth mechanism, without any change to the rule.



\subsection{Local damage under environmental and operational variability}
\label{subsec:eov_damage}

The fifth case is the one that justifies the temperature-compensation term introduced in
Eq.~\eqref{eq:eov_stiffness}. The structure is a four-degree-of-freedom shear building with floor
masses of $800$~kg and story stiffnesses $10^{6}[1.80,1.60,1.40,1.20]$~N/m at the reference
temperature, damped by a Rayleigh model at $\zeta=0.02$ (Table~\ref{tab:model_parameters}). A
broadband force is applied at the fourth floor and the response is sampled at $100$~Hz over
$180$~s. Two things change during the record. First, a benign global effect: temperature varies
about the reference of $20\,^{\circ}\mathrm{C}$ through a slow combination of harmonic and drift
terms, driving a global stiffness multiplier $\mu_K(T)$ that is constrained to the range
$0.86$ to $1.10$. Second, a genuine local fault: a $30\%$ loss of the third-story stiffness from
$t=90$~s, reducing $k_3$ from $1.40\times10^{6}$ to $0.98\times10^{6}$~N/m. The test is whether the
framework can attribute the global change to temperature and still flag the local damage.

\begin{figure}[H]
	\centering
	\begin{subfigure}[t]{0.48\textwidth}
		\centering
		\includegraphics[width=\linewidth]{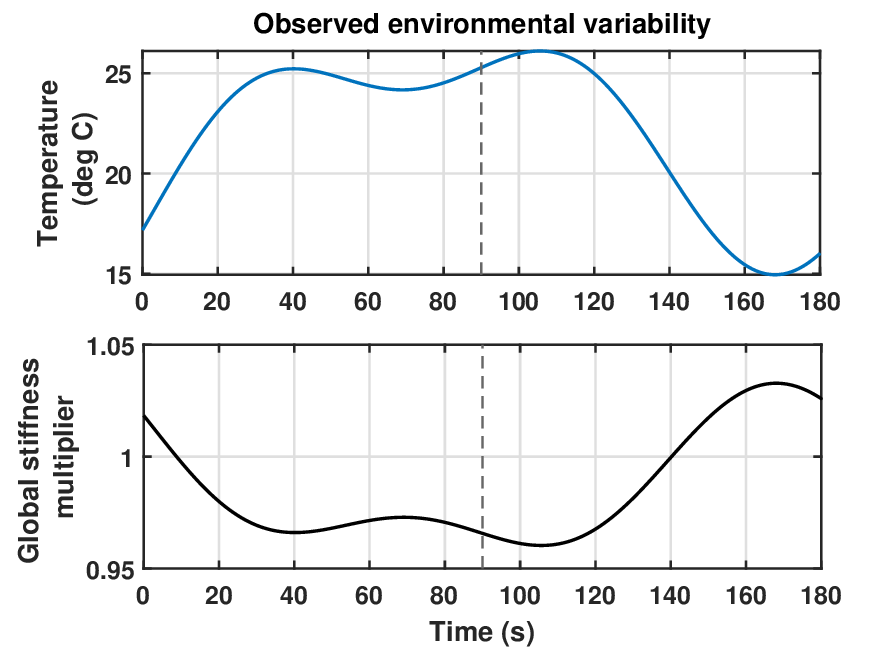}
		\caption{Temperature and $\mu_K(T)$.}
		\label{fig:case5_temperature}
	\end{subfigure}
	\hfill
	\begin{subfigure}[t]{0.48\textwidth}
		\centering
		\includegraphics[width=\linewidth]{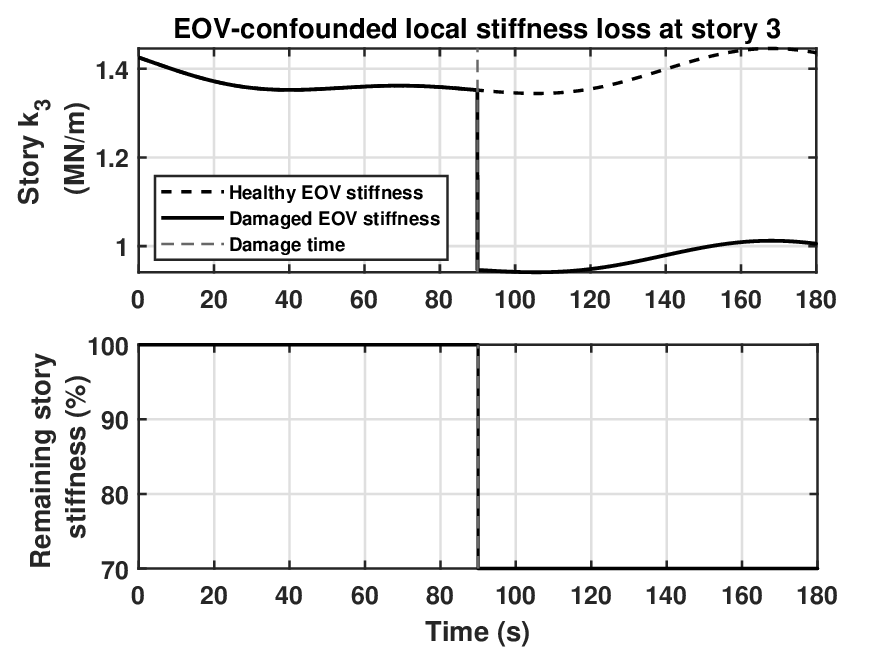}
		\caption{Confounded local stiffness.}
		\label{fig:case5_stiffness}
	\end{subfigure}
	\caption{EOV-confounded damage case, setup. (a) The observed temperature and the benign global
		stiffness multiplier it induces, which stays within a bounded range near unity. (b) The
		third-story stiffness carries both the temperature-driven global change and a $30\%$ local loss
		from $t=90$~s.}
	\label{fig:case5_setup}
\end{figure}

The two coincident changes are shown in Figure~\ref{fig:case5_setup}. The temperature varies slowly
about the reference and drives the global stiffness multiplier over a bounded range near unity
(panel \ref{fig:case5_temperature}), which is a benign, structure-wide effect. Superimposed on this is a genuine local fault:
a $30\%$ loss of the third-story stiffness from $t=90$~s (panel \ref{fig:case5_stiffness}), which the benign temperature
field cannot explain. The test for the framework is to attribute the global change to
temperature and still flag the local fault.

\begin{figure}[H]
	\centering
	\begin{subfigure}[t]{0.48\textwidth}
		\centering
		\includegraphics[width=\linewidth]{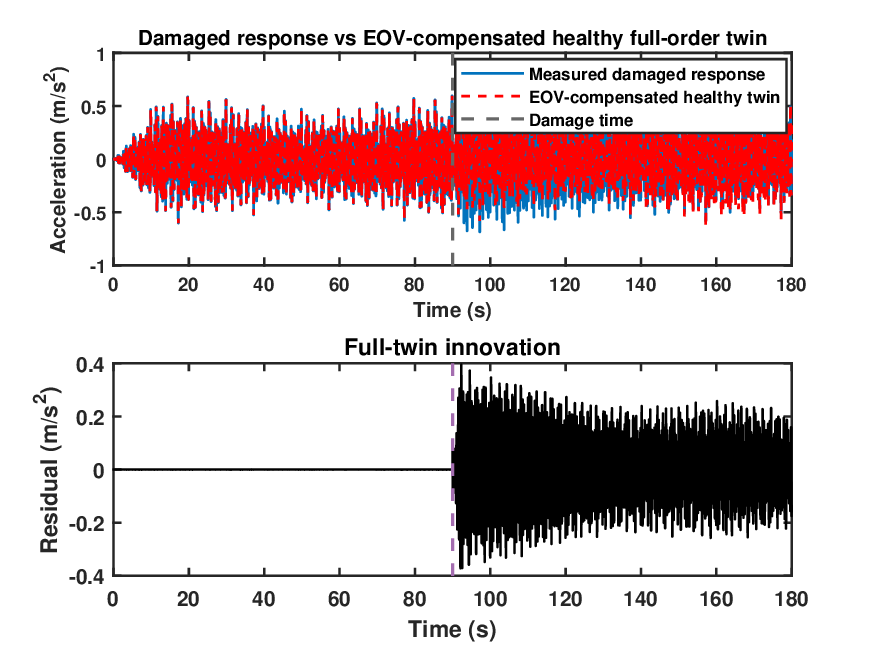}
		\caption{Compensated twin and innovation.}
		\label{fig:case5_residual}
	\end{subfigure}
	\hfill
	\begin{subfigure}[t]{0.48\textwidth}
		\centering
		\includegraphics[width=\linewidth]{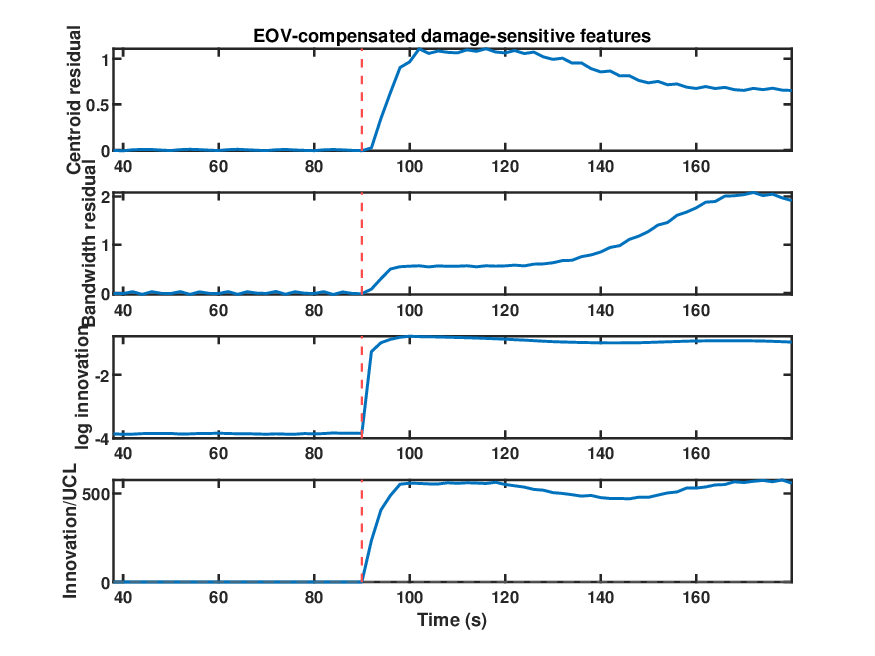}
		\caption{Compensated features.}
		\label{fig:case5_features}
	\end{subfigure}
	\caption{EOV-confounded damage case, physics evidence. (a) The temperature-compensated healthy
		twin tracks the measured response through the benign temperature variation, and the innovation
		rises only after the local stiffness loss. (b) The compensated features remain stable during
		benign variability and change at the damage time.}
	\label{fig:case5_physics}
\end{figure}

The key methodological step in this case is compensation. The healthy twin is given the same
temperature-dependent global stiffness field through Eq.~\eqref{eq:eov_stiffness}, and the ordinary
spectral features are regressed against temperature using undamaged training windows only, so that
benign variability is represented in the reference rather than treated as novelty. The effect is
shown in Figure~\ref{fig:case5_physics}. The compensated healthy twin tracks the measured response
through the temperature swings, so the innovation (panel a) stays small before $t=90$~s and rises
only after the local stiffness loss, which compensation cannot absorb. The compensated
damage-sensitive features (panel b) show the same separation: they are flat during the benign
variation and step at the damage time. 

\begin{figure}[H]
	\centering
	\begin{subfigure}[t]{0.48\textwidth}
		\centering
		\includegraphics[width=\linewidth]{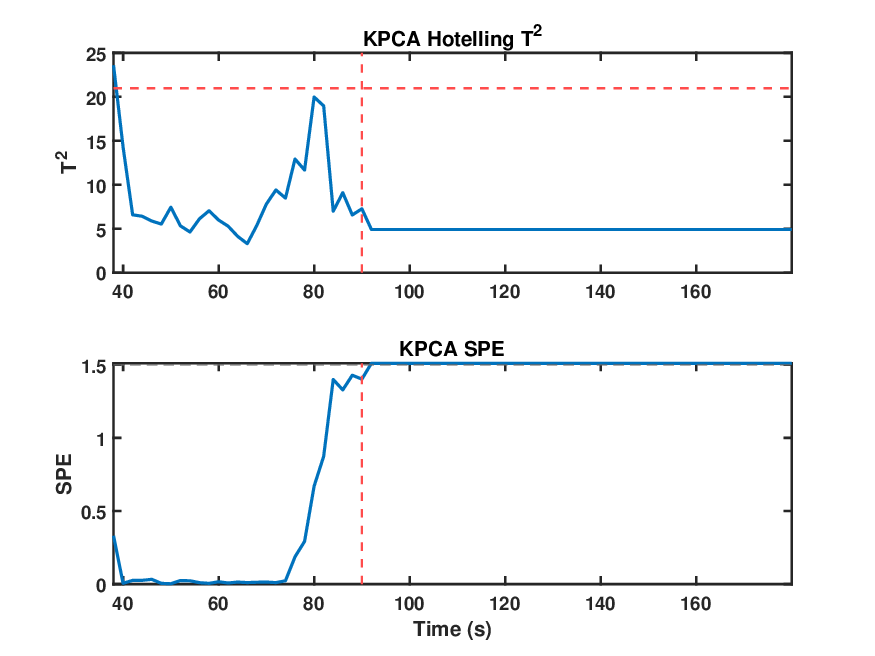}
		\caption{KPCA $T^2$ and SPE.}
		\label{fig:case5_kpca}
	\end{subfigure}
	\hfill
	\begin{subfigure}[t]{0.48\textwidth}
		\centering
		\includegraphics[width=\linewidth]{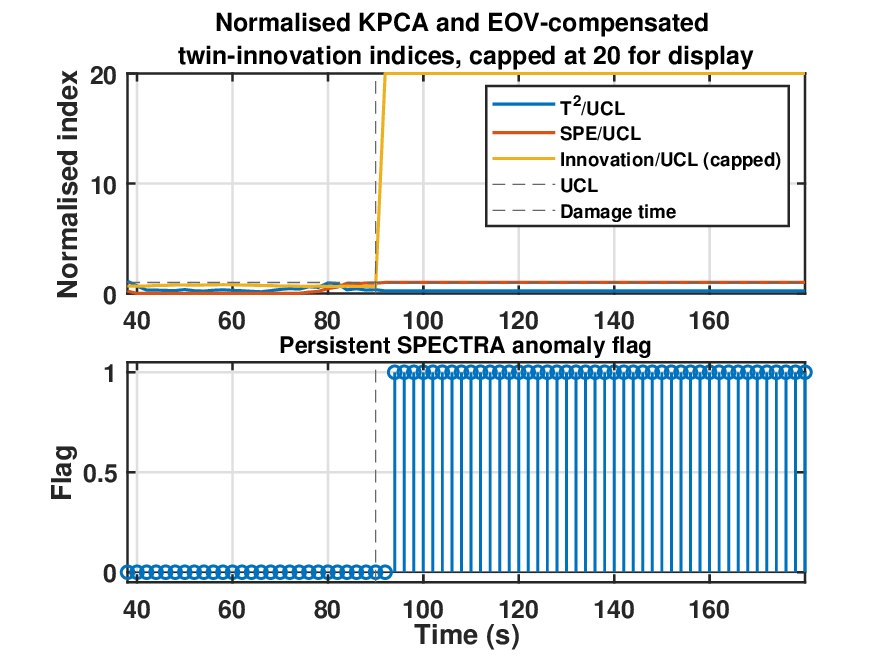}
		\caption{Normalised indices and flag.}
		\label{fig:case5_flags}
	\end{subfigure}
	\caption{EOV-confounded damage case, monitoring and decision. (a) The KPCA indices with their
		upper control limits, calibrated on undamaged windows only. (b) The normalised indices stay
		below unity through the benign temperature variation and exceed it after the local damage; the
		persistent flag produces a sustained alarm with no pre-damage false alarm.}
	\label{fig:case5_decision}
\end{figure}

The decision-level response is shown in Figure~\ref{fig:case5_decision}. The upper control limits
are calibrated only from undamaged windows, including the validation range that spans the benign
temperature variation before the known damage onset. This is a fixed reference calibration, not a
threshold adjusted after seeing the damage. The normalised innovation and SPE indices stay below
unity during the benign variation and exceed it after damage, and the persistent flag activates
after $t=90$~s with no persistent pre-damage false alarm. The contrast with the operational-only
case in Section~\ref{subsec:negative_control} is the point: there, the same benign mechanisms act
without any local fault, and the flag stays at zero.

\subsection{Local damping loss}
\label{subsec:damping_loss}

The sixth case targets a mechanism that is deliberately hard to see in the frequency domain. The
structure is again a four-degree-of-freedom shear building with uniform floor masses of $1000$~kg
and uniform story stiffnesses of $1.80\times10^{6}$~N/m, but here the damping is modelled as a
light Rayleigh background at $\zeta=0.003$ plus discrete story dampers of $2.40$~kN\,s/m
(Table~\ref{tab:model_parameters}). The response is sampled at $200$~Hz over $180$~s. Damage is an
$82\%$ loss of the second-story damper from $t=90$~s, reducing its coefficient from
$2.40$ to $0.43$~kN\,s/m, while all stiffnesses are held constant. Because natural frequencies
depend on stiffness and mass rather than damping, this fault barely moves the modal frequencies, so
it is the clearest demonstration that the twin-innovation channel provides information that the
spectral features alone do not.

\begin{figure}[H]
	\centering
	\begin{subfigure}[t]{0.48\textwidth}
		\centering
		\includegraphics[width=\linewidth]{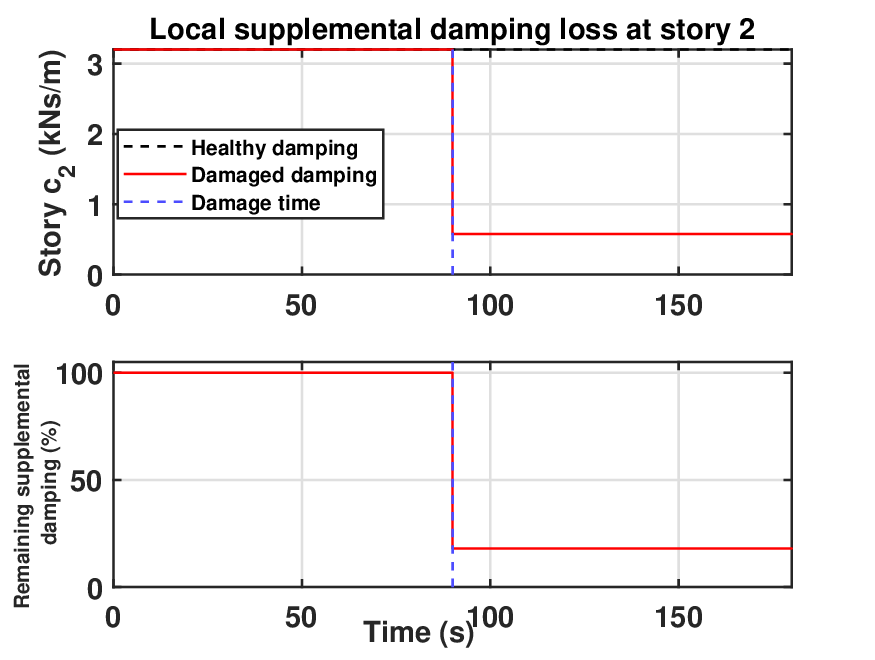}
		\caption{Damping-loss profile and drift.}
		\label{fig:case6_profile}
	\end{subfigure}
	\hfill
	\begin{subfigure}[t]{0.48\textwidth}
		\centering
		\includegraphics[width=\linewidth]{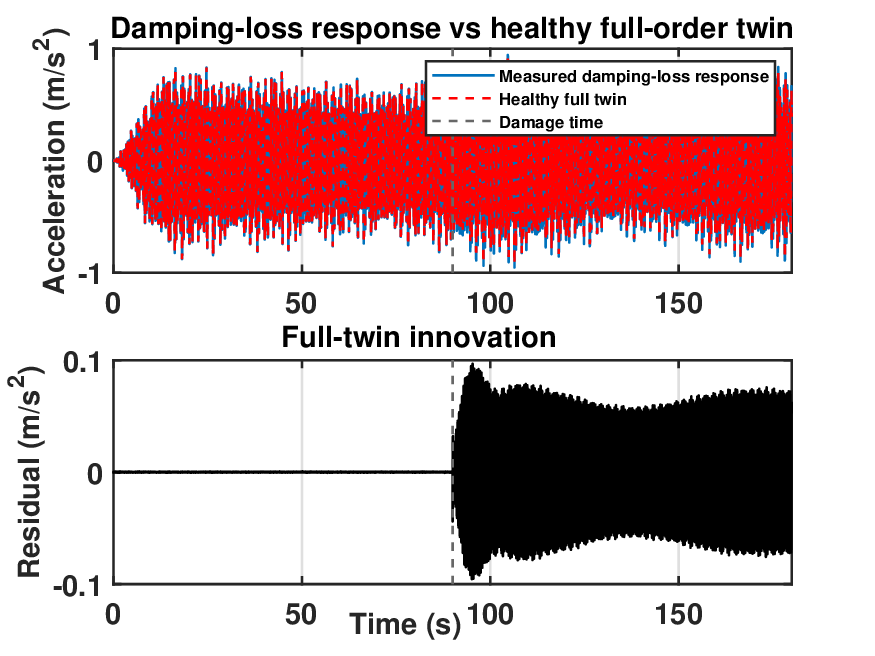}
		\caption{Twin comparison and innovation.}
		\label{fig:case6_residual}
	\end{subfigure}
	\caption{Local damping-loss case, setup and physics. (a) The effective second-story damping
		drops by $82\%$ at $t=90$~s and the corresponding story drift increases. (b) The measured
		response departs from the healthy full-order twin, and the innovation rises after the onset,
		even though the stiffness and therefore the modal frequencies are unchanged.}
	\label{fig:case6_setup}
\end{figure}

\begin{figure}[H]
	\centering
	\begin{subfigure}[t]{0.48\textwidth}
		\centering
		\includegraphics[width=\linewidth]{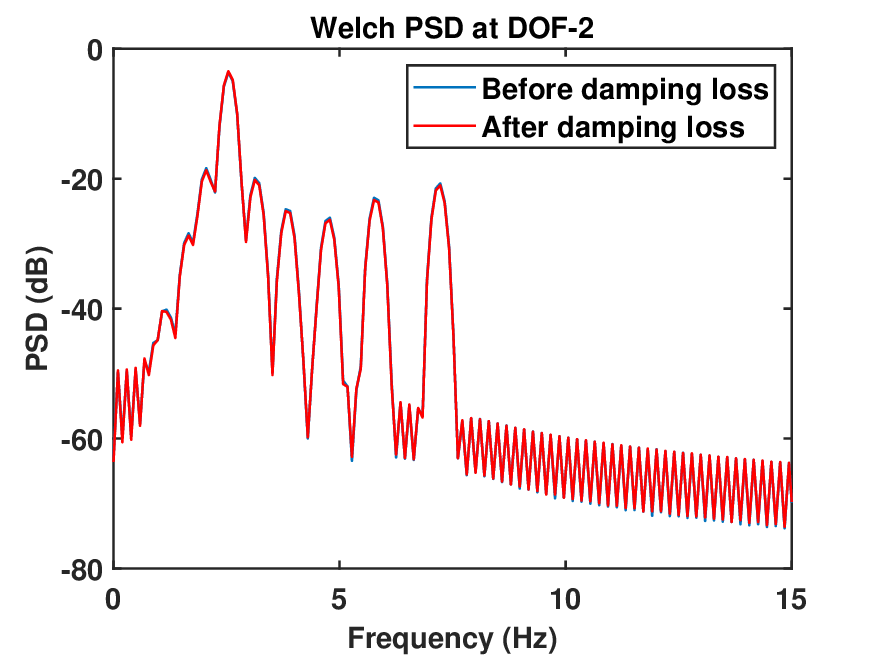}
		\caption{Welch PSD.}
		\label{fig:case6_psd}
	\end{subfigure}
	\hfill
	\begin{subfigure}[t]{0.48\textwidth}
		\centering
		\includegraphics[width=\linewidth]{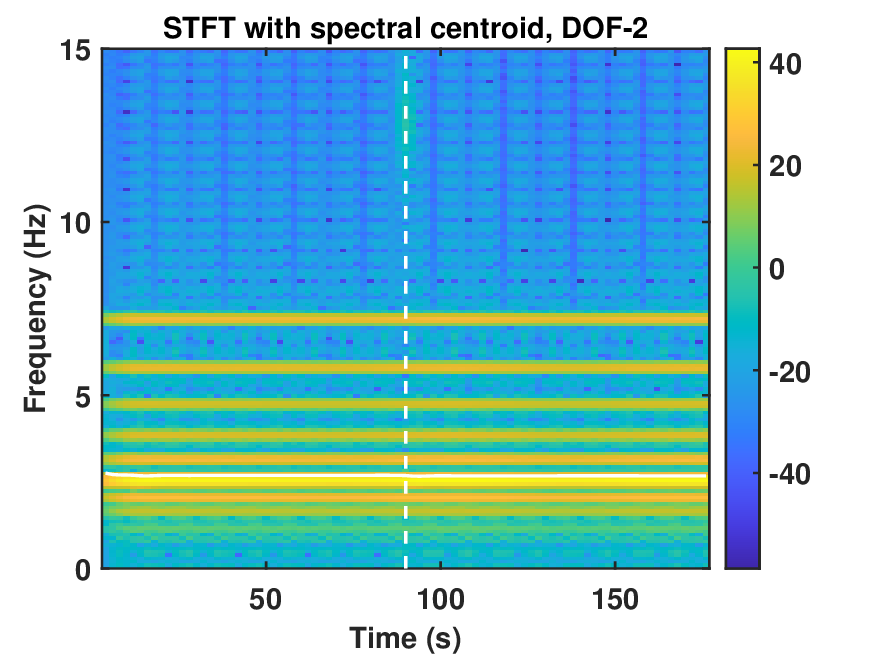}
		\caption{STFT with spectral centroid.}
		\label{fig:case6_stft}
	\end{subfigure}
	\caption{Local damping-loss case, frequency-domain evidence at DOF-2. (a) The Welch spectra
		before and after the onset differ mainly in peak amplitude and width, not location. (b) The
		spectral centroid is nearly stationary across the onset.}
	\label{fig:case6_spectra}
\end{figure}

The imposed damping-loss profile and the resulting story-2 drift are shown in
Figure~\ref{fig:case6_setup}(a), and the measured response against the healthy full-order twin,
with the innovation, in Figure~\ref{fig:case6_setup}(b). The effective damping at story 2 drops at
the onset time, and the response amplitude at that story grows because less energy is dissipated
per cycle. The healthy twin, which retains the original damping, no longer explains the measured
response. The difficulty of this case is that the fault is almost invisible in the frequency domain, as
Figure~\ref{fig:case6_spectra} shows. The Welch spectra before and after the onset (panel a) differ
mainly in peak amplitude and width, not in peak location, and the spectral centroid (panel b) is
nearly stationary across the onset. A monitoring scheme that relied on a frequency shift alone would
therefore struggle to flag this fault at all.

\begin{figure}[H]
	\centering
	\includegraphics[width=0.55\textwidth]{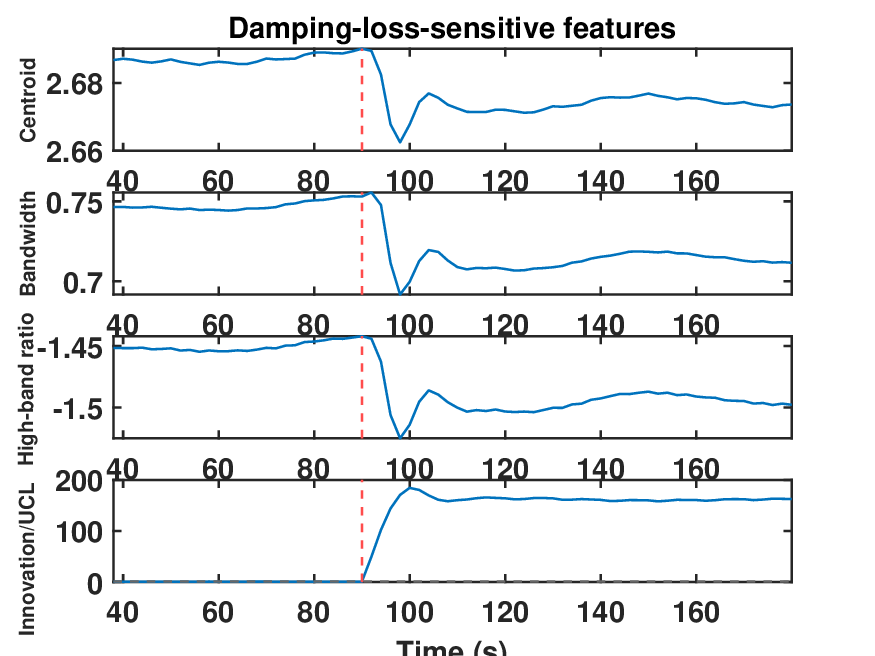}
	\caption{Local damping-loss case: damping-sensitive features. The bandwidth, peak-to-RMS ratio
		and innovation channel respond where the frequency content does not.}
	\label{fig:case6_features}
\end{figure}

\begin{figure}[H]
	\centering
	\begin{subfigure}[t]{0.48\textwidth}
		\centering
		\includegraphics[width=\linewidth]{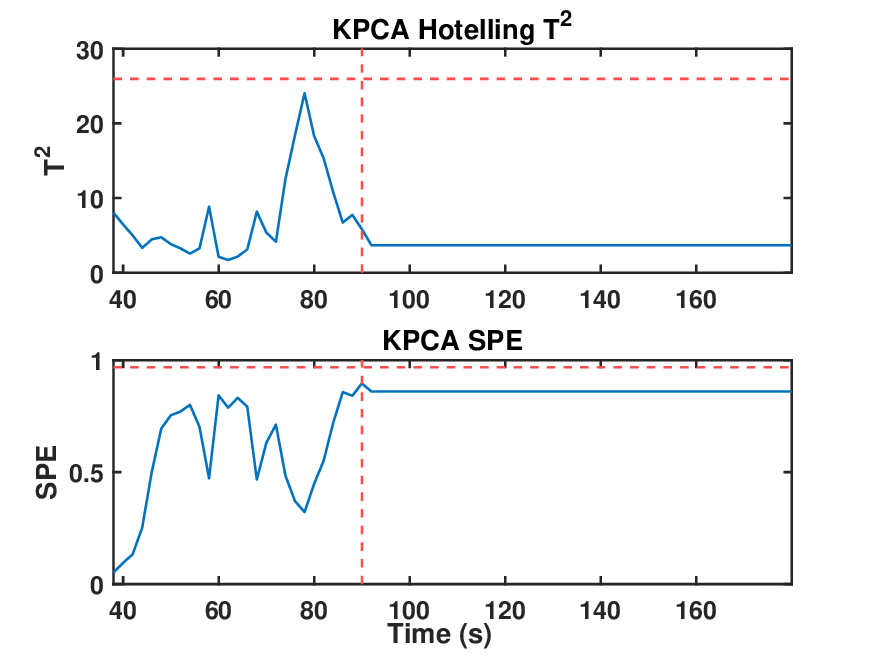}
		\caption{KPCA $T^2$ and SPE.}
		\label{fig:case6_kpca}
	\end{subfigure}
	\hfill
	\begin{subfigure}[t]{0.48\textwidth}
		\centering
		\includegraphics[width=\linewidth]{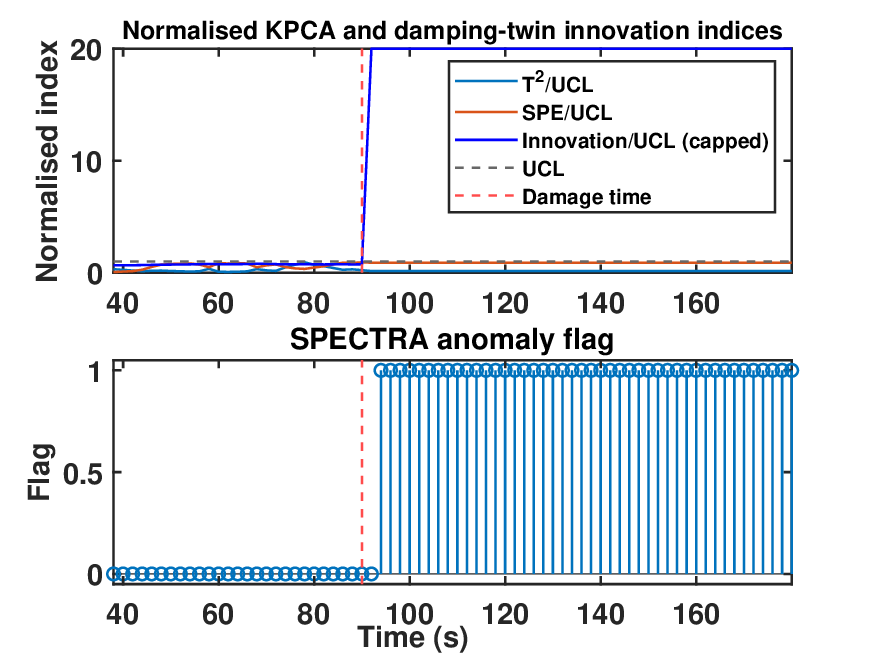}
		\caption{Normalised indices and flag.}
		\label{fig:case6_flags}
	\end{subfigure}
	\caption{Local damping-loss case, monitoring and decision. (a) The KPCA indices respond weakly as the feature manifold moves little. (b) The innovation index carries the detection and the persistent flag produces a sustained alarm after the onset, with no pre-damage false alarm.}
	\label{fig:case6_decision}
\end{figure}

The damping-sensitive features tell a different story, shown in Figure~\ref{fig:case6_features}.
While the centroid barely moves, the bandwidth, the peak-to-RMS ratio and, most strongly, the
innovation channel all respond to the loss of damping, because less energy is dissipated per cycle
and the healthy twin no longer matches the measured decay. The decision-level response is shown in Figure~\ref{fig:case6_decision}. The KPCA $T^2$ and SPE
indices respond only weakly, consistent with the small movement of the feature manifold, while the
innovation index rises clearly after the onset. This is the case in which the innovation channel is
decisive: it carries the alarm when the KPCA indices remain near their limits. The persistent flag
activates after $t=90$~s with no persistent pre-damage false alarm. The case therefore supports the
use of a combined decision structure rather than any single damage-sensitive index, since a
frequency-based index alone would have been near-silent here.

\subsection{Operational-only negative control}
\label{subsec:negative_control}

The seventh case is the most important for establishing specificity, and it is a negative control:
the structure remains healthy throughout, but its operating conditions change. Both the temperature
and the excitation regime vary during the record, with no local stiffness or damping fault
introduced at any time. The purpose is to confirm that the framework does not convert benign
operational and environmental change into a damage alarm. A monitoring method that flags this case
would be unusable in the field, where loading and environment are never stationary.

\begin{figure}[h]
	\centering
	
	\begin{minipage}[t]{0.48\textwidth}
		\centering
		\includegraphics[width=0.95\linewidth]{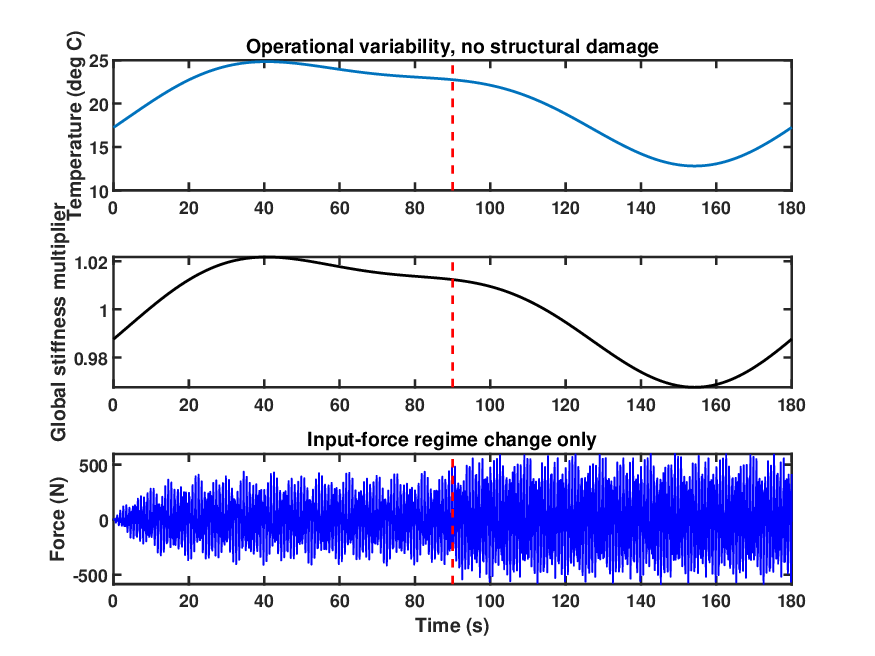}
		\caption{Operational-only negative-control case showing the imposed variability. 
			The temperature cycle induces a global stiffness multiplier that remains close to unity, 
			while the input-force regime changes at $t=90$~s. No structural fault is introduced.}
		\label{fig:case7_setup}
	\end{minipage}
	\hfill
	\begin{minipage}[t]{0.48\textwidth}
		\centering
		\includegraphics[width=0.90\linewidth]{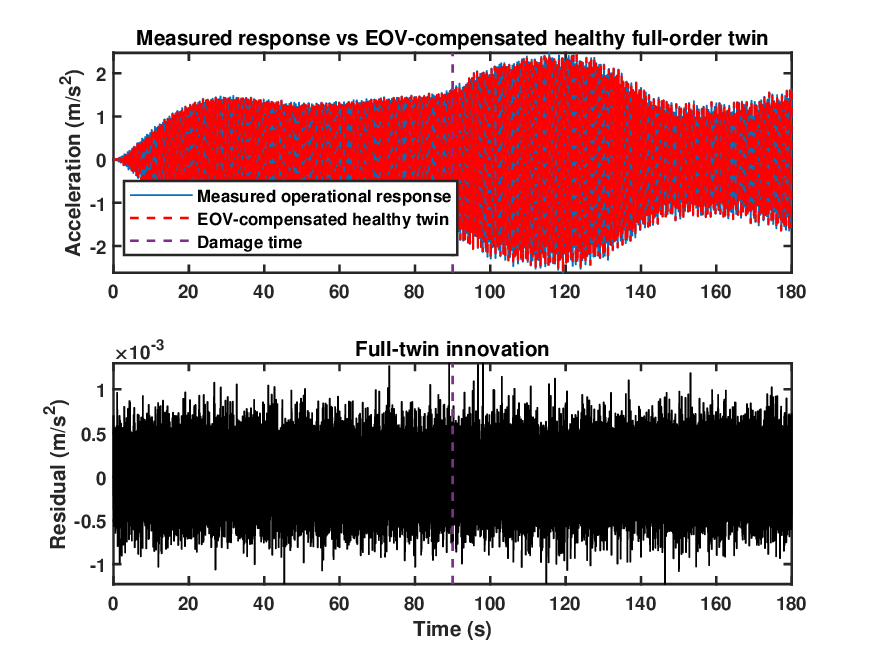}
		\caption{Measured response for the operational-only negative-control case compared with the temperature-compensated healthy full-order twin. 
			The innovation RMS remains essentially unchanged across the operational transition, 
			from $3.0165\times10^{-4}$ before the transition to $3.0194\times10^{-4}$ after, confirming that the healthy twin explains the response.}
		\label{fig:case7_innovation}
	\end{minipage}
	
\end{figure}

The operational variability imposed in this case is shown in Figure~\ref{fig:case7_setup}: the
temperature follows a slow cycle, the induced global stiffness multiplier stays within a few per
cent of unity, and the input-force regime changes at $t=90$~s. No structural fault is introduced at
any point, so any change in the measured response is benign by construction. The decisive evidence is the twin innovation, shown in Figure~\ref{fig:case7_innovation}. The
temperature-compensated healthy twin continues to track the measured response across the operational
transition, and the innovation root-mean-square value is $3.0165\times10^{-4}$ before the transition
and $3.0194\times10^{-4}$ after it, a change of less than one part in a thousand. In physics-space
terms the healthy twin still explains the measured response, even though the operating regime has
changed. This near-equality is the property the persistent decision rule depends on.

\begin{figure}[H]
	\centering
	\includegraphics[width=0.65\textwidth]{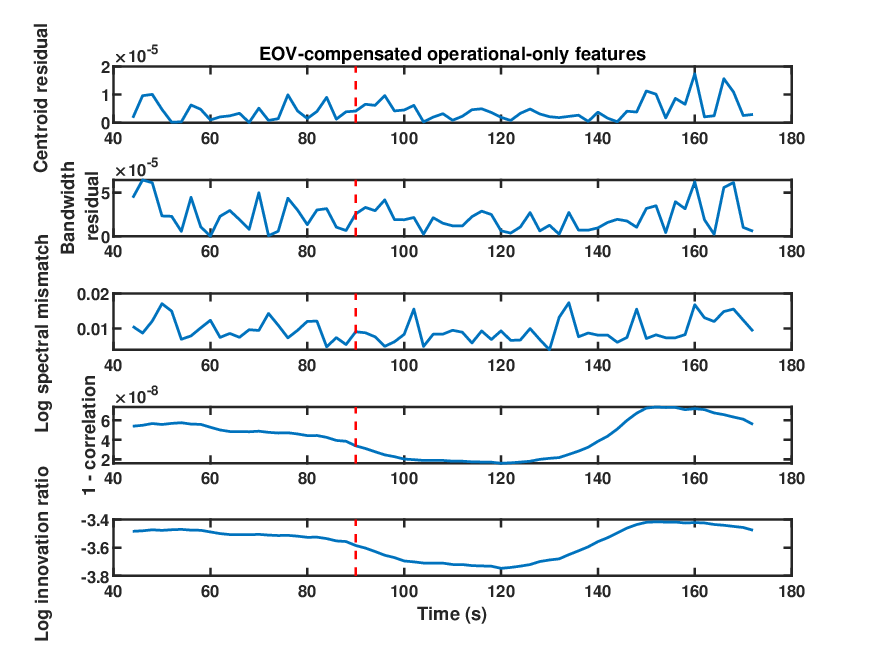}
	\caption{Operational-only negative-control case: EOV-compensated operational-only features. The
		feature space moves under the new operating regime, with no structural fault present.}
	\label{fig:case7_features}
\end{figure}

\begin{figure}[H]
	\centering
	\begin{subfigure}[t]{0.48\textwidth}
		\centering
		\includegraphics[width=\linewidth]{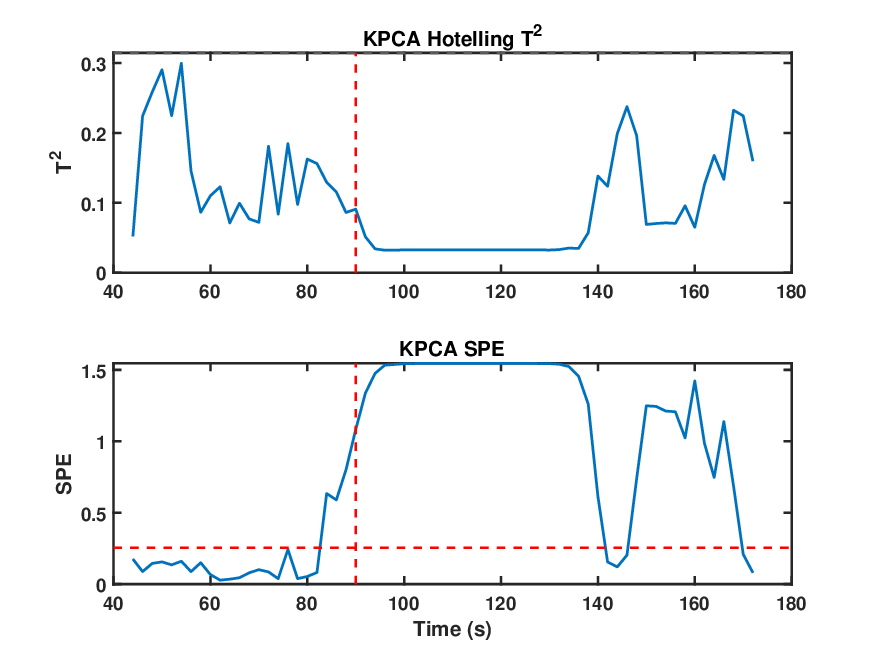}
		\caption{KPCA $T^2$ and SPE.}
		\label{fig:case7_kpca}
	\end{subfigure}
	\hfill
	\begin{subfigure}[t]{0.48\textwidth}
		\centering
		\includegraphics[width=\linewidth]{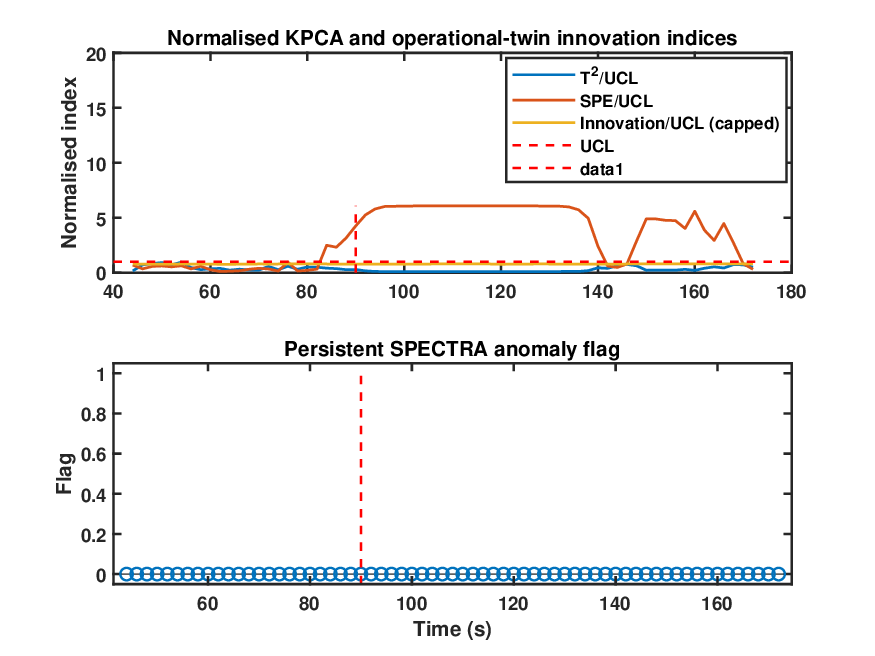}
		\caption{Normalised indices and persistent flag.}
		\label{fig:case7_flags}
	\end{subfigure}
	\caption{Operational-only negative-control case, monitoring and decision. (a) The KPCA SPE index
		rises above its limit for a sustained period after the operational transition. (b) The normalised
		indices show SPE exceeding unity, yet the innovation channel stays within its limit and the
		persistent flag remains zero throughout.}
	\label{fig:case7_decision}
\end{figure}

The feature space, by contrast, is genuinely disturbed by the regime change, as
Figure~\ref{fig:case7_features} shows. The compensated features move at the operational transition,
which is exactly the behaviour that would mislead a novelty-only detector. The point of the case is
that this feature-space movement is not, on its own, allowed to raise a structural alarm. The monitoring indices and the final decision are shown in Figure~\ref{fig:case7_decision}. The KPCA SPE index in panel~(a) rises above its upper control limit and stays there for an extended period. This rise occurs because the feature distribution changes when the input regime changes. The normalised indices and the persistent flag in panel~(b) show the consequence: SPE exceeds unity, yet the flag remains at zero throughout because the innovation channel that gates the alarm in Eq.~\eqref{eq:raw_alarm} stays within its limit. This is the central result of the negative control:
the SPE channel alone would raise a sustained false alarm here, and it is the physics-space
innovation, not the statistical novelty, that keeps the decision inactive.

\begin{table}[H]
	\centering
	\caption{Summary of SPECTRA benchmark outcomes. The two no-damage cases are interpreted as specificity tests rather than damage-detection cases.}
	\label{tab:benchmark_outcomes}
	\small
	\resizebox{\textwidth}{!}{%
		\begin{tabular}{lllllll}
			\hline
			Case & Structural condition & Pre-event innovation RMS & Post-event innovation RMS & Pre-event flags & Post-event flags & Decision outcome \\
			\hline
			SPECTRA 01 & Duffing drift, no damage & -- & -- & 0 & 0 & No persistent false alarm \\
			SPECTRA 02 & 30\% step stiffness loss, story 2 & \(4.6442\times10^{-4}\) & \(1.5959\times10^{-1}\) & 0 & 43 & Detected, \(2.00~\mathrm{s}\) delay \\
			SPECTRA 03 & Gradual 30\% stiffness loss, story 2 & \(1.1068\times10^{-4}\) & \(3.5965\times10^{-1}\) & 0 & 44 & Detected, \(3.99~\mathrm{s}\) delay \\
			SPECTRA 04 & Bilinear breathing stiffness, story 2 & \(1.9546\times10^{-4}\) & \(3.5709\times10^{-1}\) & 0 & 44 & Detected, \(3.99~\mathrm{s}\) delay \\
			SPECTRA 05 & EOV-confounded 30\% stiffness loss, story 3 & \(1.3838\times10^{-4}\) & \(1.1430\times10^{-1}\) & 0 & 44 & Detected, \(3.99~\mathrm{s}\) delay \\
			SPECTRA 06 & 82\% damping loss, story 2 & \(4.2833\times10^{-5}\) & \(1.0571\times10^{-2}\) & 0 & 43 & Detected, \(2.00~\mathrm{s}\) delay \\
			SPECTRA 07 & Operational-only change, no damage & \(3.0165\times10^{-4}\) & \(3.0194\times10^{-4}\) & 0 & 0 & No persistent false alarm \\
			\hline
	\end{tabular}}
\end{table}

Table~\ref{tab:benchmark_outcomes} shows that the persistent decision rule gives zero pre-event persistent false flags in all damage cases and zero persistent false alarms in both no-damage cases. The result is not obtained by one monitoring index alone. In the stiffness-loss and breathing-stiffness cases, SPE and innovation both provide strong evidence of damage. In the damping-loss case, the innovation channel is the dominant evidence source. In the operational-only negative-control case, SPE exceeds its limit, but the innovation channel remains below its control limit and the persistent damage flag stays at zero. This confirms that SPECTRA separates feature-space novelty from structural anomaly.

\section{Discussion}
\label{sec:discussion}

The numerical studies show that SPECTRA should be interpreted as a physics-informed decision framework rather than as a single-index novelty detector. Across the seven benchmark cases, the final decision is governed by the relationship between feature-space novelty, healthy-twin disagreement and persistence. This distinction is important because the same statistical index can have different meanings in different physical settings. In the stiffness-loss and breathing-stiffness cases, SPE and the innovation channel both provide strong evidence of structural change. In the damping-loss case, the innovation channel is the dominant source of evidence because the frequency-sensitive features change weakly. In the operational-only negative-control case, SPE rises above its limit under benign operational change, but the innovation channel remains below its control limit and the persistent flag stays inactive. The discussion below therefore focuses on three issues: how the three monitoring channels should be interpreted, how SPECTRA is positioned relative to existing monitoring approaches, and what limitations bound the present numerical evidence.

\subsection{Role of the three monitoring channels}
The benchmark suite was designed so that no single monitoring channel carries every case, and the
results support this design. In the abrupt and gradual stiffness-loss cases, the squared prediction
error and the innovation channel both separate clearly from the healthy state, since the softened
structure leaves the healthy feature manifold and departs from the healthy twin at the same time. In
the breathing-stiffness case, the response is generated by a mechanism the linear twin cannot
represent, so the innovation and the squared prediction error carry the detection while the
Hotelling statistic responds only weakly. The damping-loss case is the most instructive: the modal
frequencies barely move, the kernel indices stay near their limits, and the innovation channel alone
provides a decisive signal. In the environmental and operational variability case, temperature
compensation keeps the innovation small during benign global change and allows it to rise only after
the local fault, so the framework attributes the global change to temperature while still flagging
the damage.

The negative control ties these observations together. When only the operating regime changes, the
squared prediction error rises above its limit and remains there for an extended period, yet the
persistent flag stays at zero. The reason is that the innovation channel, which gates the raw alarm
in Eq.~\eqref{eq:raw_alarm}, stays within its limit throughout, since the compensated twin continues
to explain the measured response. Across all seven cases the persistent flag therefore follows the
innovation channel: it latches when the measured response is no longer consistent with the healthy
twin, and it stays inactive when the response changes for benign reasons only. This gated structure,
rather than the kernel indices in isolation, is what gives the framework its specificity.

\subsection{Relation to existing approaches}
Table~\ref{tab:comparison} positions SPECTRA against the main families of monitoring methods.
Data-driven novelty detection, including kernel principal component analysis, is sensitive to change
but cannot on its own separate benign operational variability from structural deterioration
\cite{worden2007machine,sohn2007eov,scholkopf1998kpca}. Modal and frequency-based methods are
physically interpretable but lose sensitivity for damping-dominated and non-smooth faults, and they
are known to be confounded by environmental effects \cite{peeters2001z24}. Recursive subspace and eigen-perturbation methods deliver genuine real-time updating, but they remain data-driven and do not provide an explicit physical reference \cite{bhowmik2020rcca_mssp}.
Physics-based digital twins supply such a reference, yet the emphasis has largely been on life
prediction and model updating rather than on online specificity under operational variability
\cite{tuegel2011digital,torzoni2024digital}. SPECTRA combines a physics-informed twin
with kernel novelty indices under a gated persistent rule, and the negative-control result shows that
this combination suppresses the operational false alarms that a novelty-only detector would raise.


\begin{table}[htbp]
	\centering
	\caption{Qualitative comparison of monitoring approach families against the capabilities exercised in the present benchmark suite. ``Yes'' indicates that the capability is directly addressed by the approach, ``Partial'' indicates that it can be addressed with additional modelling or assumptions, and ``No'' indicates that it is not normally addressed by the approach in its standard form.}
	\label{tab:comparison}
	\resizebox{\textwidth}{!}{%
		\begin{tabular}{lccccccc}
			\hline
			Approach & Abrupt / gradual & Non-smooth & Damping- & EOV & Operational & Physics & Real- \\
			& stiffness loss   & breathing  & only     & robust & false-alarm & reference & time \\
			\hline
			Data-driven novelty / KPCA \cite{worden2007machine,scholkopf1998kpca} &
			Yes & Partial & Partial & No & No & No & Partial \\
			Modal / frequency-based \cite{peeters2001z24} &
			Yes & Partial & No & Partial & Partial & Partial & Partial \\
			Recursive subspace / eigen-perturbation \cite{bhowmik2019foep,bhowmik2024rsd} &
			Yes & Partial & Partial & Partial & Partial & No & Yes \\
			Physics-based digital twin \cite{tuegel2011digital,torzoni2024digital} &
			Yes & Partial & Partial & Partial & Partial & Yes & Partial \\
			SPECTRA, present benchmark suite &
			Demonstrated & Demonstrated & Demonstrated & Demonstrated & Demonstrated & Yes & Yes \\
			\hline
	\end{tabular}}
\end{table}

\subsection{Limitations}
Three limitations should be stated plainly, since they bound the claims that the present results
support. First, the healthy twin is given the exact reference mass, damping and stiffness matrices and
the same excitation as the monitored system, with a deterministic noise floor added only to the
measured record. This isolates the decision logic from twin-estimation error, which is the intended
purpose of a controlled benchmark, but it also means that the absolute magnitudes of the innovation,
often several orders of magnitude between the healthy and damaged states, are optimistic. In a field
deployment the twin must be fitted and updated from data, and the claim that transfers is the gated
decision logic rather than the absolute innovation scale. Second, every damage onset is fixed at
$t=90$~s and the excitation and environment are controlled, which is appropriate for a like-for-like
comparison across cases but does not represent the timing and non-stationarity of field loading.
Third, the study is confined to numerical shear-building models with single-degree-of-freedom
monitoring and fixed control limits calibrated on healthy windows; performance under long-horizon
environmental drift, multi-sensor fusion and modelling error remains to be established. These points
motivate the validation programme set out in Section~\ref{sec:conclusions}.

\section{Conclusions and future work}
\label{sec:conclusions}

This paper presented SPECTRA, a physics-informed eigen-compressed digital twin framework for
real-time structural anomaly inference. The framework integrates a healthy physics-informed twin, an
eigen-compressed dynamic representation, residual-augmented spectral features, kernel principal
component analysis and an innovation-gated persistent decision rule into a single inference pipeline.
The central idea is that a structural anomaly is declared not from statistical novelty alone, but
from persistent disagreement between the measured response and the healthy twin, corroborated by
kernel novelty.

Across seven controlled benchmarks the framework detected abrupt, gradual, nonlinear breathing and
damping-dominated damage with finite detection delay, handled a local fault confounded by
temperature-driven variability, and produced no persistent alarm in an operational-only
negative-control case. The most informative result is the negative control: the kernel squared
prediction error rose above its limit when the operating regime changed, yet the persistent flag
stayed at zero because the physics-space innovation that gates the decision remained within its
limit. Specificity in SPECTRA therefore comes from the physics-informed innovation rather than from
the statistical indices in isolation, and this is the property that distinguishes it from
novelty-only monitoring.

The limitations identified in Section~\ref{sec:discussion} set a clear programme of further work. The immediate priority is experimental and field validation, replacing the idealised exact twin with a model fitted and recursively updated from measured data. Recursive simultaneous diagonalisation and related online modal-identification procedures provide one route for updating modal information as data arrive \cite{bhowmik2024rsd}. A second route is to couple the present decision rule with adaptive decomposition and feature-extraction procedures, including multivariate SSA-type monitoring frameworks \cite{mehulkumar2026famessa}. Field and laboratory validation should then be carried out on instrumented structures where measured response, excitation uncertainty and environmental variability are all present. More broadly, the implementation should follow established principles for real-time SHM of vibrating systems \cite{bhowmik2022book}. A further extension is to connect the SPECTRA decision layer with autonomous agentic SHM frameworks, where perception, inference, action and reflection are linked through adaptive monitoring policies. This would allow the persistent anomaly decision to inform not only detection, but also sensing strategy, model updating and inspection prioritisation \cite{sharma2026autonomous}. Embedding the gated inference within digital-twin decision-support workflows \cite{torzoni2024digital} would then allow the framework to support defensible engineering decisions on operating infrastructure, rather than only reporting statistical novelty.

\bibliographystyle{unsrt}
\bibliography{references}  






\end{document}